\documentclass[10pt,aps,prd,twocolumn,showpacs,superscriptaddress,longbibliography,nofootinbib]{revtex4-2}
\usepackage{mathrsfs,amsmath,amsthm,latexsym,amssymb,amsfonts,epsfig,cancel,enumerate,graphicx,txfonts,diagbox}
\usepackage{multirow}
\usepackage[T1]{fontenc}
\usepackage[utf8]{inputenc}
\usepackage[colorlinks=true,linkcolor=blue,citecolor=blue,urlcolor=blue]{hyperref}
\usepackage{orcidlink}

\usepackage{xcolor}
\usepackage{booktabs}
\usepackage{graphicx}
\usepackage{multirow}

\definecolor{navy}{RGB}{0,0,150}

\usepackage{appendix}

\allowdisplaybreaks

\newcommand{\RGU}{Department of Physics, The Assam Royal Global University, Guwahati-781035, Assam, India}

\begin{document}

\title{Dyonic ModMax Black Holes in Kalb-Ramond gravity with a Cloud of Strings as Source
}

\author{Faizuddin Ahmed\orcidlink{0000-0003-2196-9622}}
\email{faizuddinahmed15@gmail.com}
\affiliation{\RGU}

\author{Edilberto O. Silva\orcidlink{0000-0002-0297-5747}}
\email{edilberto.silva@ufma.br}
\affiliation{Programa de P\'os-Gradua\c c\~ao em F\'{\i}sica \& Coordena\c c\~ao do Curso de F\'{\i}sica -- Bacharelado, Universidade Federal do Maranh\~{a}o, 65085-580 S\~{a}o Lu\'{\i}s, Maranh\~{a}o, Brazil}

\begin{abstract}
We investigate the geodesic structure, shadow, thermodynamics, and Hawking radiation from a dyonic ModMax black hole in Kalb-Ramond gravity with a cloud of strings. The combined presence of ModMax nonlinear electrodynamics, the
Lorentz-violating Kalb-Ramond background, and the string cloud breaks asymptotic flatness and introduces a global conical deficit that modifies all observables through a single geometric prefactor. We derive analytic expressions for the photon sphere, critical impact parameter, and shadow radius, and show that the shadow size depends on both the non-flat asymptotics and the ModMax screening of the dyonic charge. For massive test particles, we determine the innermost stable circular orbit and the accretion efficiency as functions of all model parameters. We also establish the first law of black hole thermodynamics and the generalized Smarr relation for this solution, identify a Hawking-Page-type phase transition in the specific heat, and compute the spectral energy emission rate, which we show is directly governed by the shadow radius in the geometric-optics limit. Our results demonstrate that the interplay of these three ingredients produces a phenomenology observationally distinguishable from standard Reissner-Nordstr\"{o}m black holes.
\end{abstract}

\maketitle


\section{Introduction}\label{sec:intro}

Black holes (BHs) are among the most fundamental objects in theoretical physics, exact solutions of Einstein's field equations in which gravity is strong enough to prevent even light from escaping. Since the first direct imaging of the supermassive BH shadow in M87* by the Event Horizon Telescope (EHT)
Collaboration~\cite{EHT2019L1,EHT2019L4,EHT2019L6} and the subsequent resolved image of Sgr~A* at the centre of the Milky Way~\cite{EHT2022L12,EHT2022L17}, BH physics has entered a new era of precision observational astrophysics. These images, combined with multi-messenger analyses, have placed competitive constraints on the parameters of a wide variety of BH models and theories of gravity beyond general relativity~\cite{Vagnozzi2023}. In parallel, theoretical advances motivate the study of BH solutions that incorporate additional fields, symmetry-breaking mechanisms, or nonlinear electromagnetic couplings~\cite{Vagnozzi2023,EHT2019L6}.

Nonlinear electrodynamics (NLE) provides a natural and well-motivated framework for such extensions. Classical NLE models begin with the Born-Infeld theory~\cite{BornInfeld1934}, which was designed to remove the electromagnetic self-energy divergence of the electron and arises as the effective world-volume action of D-branes in string theory. Despite their successes, both Born-Infeld and Euler-Heisenberg theories break the conformal invariance of source-free Maxwell electrodynamics in four dimensions. The question of whether a nonlinear, interacting generalization of Maxwell's equations can simultaneously preserve both conformal invariance and the SO(2) electromagnetic duality of the linear theory was unanswered for many years.

In 2020, Bandos et al. proved that the only nonlinear extensions of source-free Maxwell's equations that preserve both conformal invariance and SO(2) electromagnetic duality form a one-parameter family, now universally known as Modified Maxwell (ModMax) theory~\cite{Bandos2020,Kosyakov2020}. The ModMax Lagrangian depends on a single dimensionless parameter $\gamma \geq 0$, reducing to standard Maxwell electrodynamics for $\gamma = 0$. For any $\gamma > 0$ the field equations are nonlinear, yet the theory still admits exact light-velocity plane-wave solutions of arbitrary polarization and predicts vacuum birefringence in electromagnetic backgrounds. Subsequent investigations have explored the supersymmetric extensions of ModMax~\cite{Bandos2021Susy}, its connection to $T\bar{T}$-like deformations~\cite{Babaei2022}, its particle-physics aspects~\cite{Lechner2022}, and generalized versions of the model~\cite{Kruglov2022}.

When coupled to General Relativity, ModMax theory immediately yields new BH solutions. Flores-Alfonso, Gonz\'{a}lez-Morales, Linares, and Maceda~\cite{FloresAlfonso2021} derived the first static, spherically symmetric charged BH solutions in Einstein-ModMax gravity, finding a Reissner-Nordstr\"{o}m-like geometry in which the parameter $\gamma$ acts as a screening factor that exponentially reduces the
effective electromagnetic charge; in particular, the extremal condition can be satisfied for BH masses smaller than the bare charge value.
Dyonic extensions carrying both electric and magnetic charges were also obtained in that context, and their shadow morphology, gravitational lensing, quasinormal modes, and greybody bounds were studied in Ref.~\cite{Pantig2022} using EHT data for M87* and Sgr~A* to constrain the screening parameter. Further investigations have addressed accelerating ModMax BHs~\cite{Barrientos2022}, Taub-NUT solutions in conformal  electrodynamics~\cite{BallonBordo2021}, BH thermodynamics with NLE fields~\cite{Bokulic2021}, and charged BHs in $F(R)$-ModMax
gravity~\cite{EslamPanah2024}. Several recent investigations of ModMax black hole solutions, including cases with and without topological defects and cosmological constant, have been reported in \cite{Ahmed2026a, Ahmed2026e, Ahmed2026c, Ahmed2026d, Bokulic2025}.

An independent and observationally motivated extension of general relativity is provided by Kalb-Ramond (KR) gravity. The rank-2 antisymmetric tensor field introduced by Kalb and Ramond~\cite{KalbRamond1974} arises naturally in the low-energy limit of string theory and, when it acquires a nonzero vacuum expectation value (VEV), the non-minimal coupling between this VEV and the Ricci tensor spontaneously breaks local Lorentz symmetry~\cite{KosteleckySamuel1989}. Yang et al.~\cite{Yang2023} derived exact static, spherically symmetric BH solutions in KR gravity, obtaining a Schwarzschild-like geometry deformed by the dimensionless Lorentz-violating parameter $\ell$; the horizon radius, thermodynamics, and Solar-System tests all exhibit a characteristic sensitivity to $\ell$. Earlier, Lessa et al.~\cite{Lessa2020} had obtained a modified BH with a
Reissner-Nordstr\"{o}m-like two-horizon structure generated purely by the KR VEV, in the absence of any electric charge. The gravitational-lensing, shadow-radius, and geodesic-precession phenomenology of KR BHs has been studied extensively~\cite{Junior2024,Filho2024,Liu2024constraints}, and
the Lorentz-violating parameter has been constrained via EHT shadow data for both M87* and Sgr~A*, as well as via Gravity Probe~B geodetic precession measurements~\cite{Liu2024constraints}.

A complementary class of exotic matter that modifies the BH metric in a
qualitatively distinct way is the cloud of strings (CoS) introduced by
Letelier~\cite{Letelier1979}. In his seminal work, Letelier showed that a spherically symmetric cloud of fundamental strings acting as a gravitational source yields a lapse function containing an asymptotically constant factor $(1-\alpha)$, where $\alpha \in (0,1)$ is the string-cloud parameter. This conical deficit induces a departure from asymptotic flatness and alters both the
horizon structure and the thermodynamic properties of the BH. The Letelier spacetime has served as a versatile testbed for exploring how macroscopic string distributions influence photon orbits, BH shadows, quasinormal modes, and weak gravitational lensing in numerous modified-gravity contexts~\cite{Toledo2019,Chabab2020,Belhaj2022}.

Combining the three independently well-motivated ingredients, ModMax nonlinear electrodynamics, Kalb-Ramond gravity, and a cloud of strings, is a natural and largely unexplored step. The simultaneous presence of the ModMax nonlinearity (parametrized by $\gamma$), spontaneous Lorentz symmetry breaking (parametrized by $\ell$), and a distributional string source (parametrized by $\alpha$) considerably enriches the spacetime geometry. The metric function departs from unity at infinity due to the combined conical and Lorentz-violating effects, while the electromagnetic sector departs from the standard RN structure through the exponential screening factor $e^{-\gamma}$. Previous work by the present authors~\cite{Ahmed2026a,Ahmed2026b} investigated KR-ModMax BHs in the absence of a string cloud and found significant modifications to the photon sphere, shadow, and thermodynamic properties. The present paper extends that programme to the dyonic case with a cloud of strings.

In this work, we derive and analyze a static, spherically symmetric dyonic ModMax BH carrying both electric charge $Q_e$ and magnetic charge $Q_m$ in Kalb-Ramond gravity with a cloud of strings as the matter source. The metric function takes the form of Eq.~\eqref{aa2}, which simultaneously generalizes the Reissner-Nordstr\"{o}m geometry by the combined action of the string-cloud parameter $\alpha$, the KR Lorentz-violating parameter $\ell$, and the ModMax nonlinearity $\gamma$. We carry out a systematic study of the photon sphere radius $r_s$ and critical impact parameter $\beta_c$, the BH shadow radius $R_{\rm sh}$ for both finite and distant observers, the effective potentials and radial force for photons, and the effective potential, specific angular momentum $\mathcal{L}_{\rm sp}$, and specific energy $\mathcal{E}_{\rm sp}$ for massive neutral test particles, including the location of the innermost stable circular orbit (ISCO). We further derive the Hawking temperature, entropy, Smarr relation, specific heat capacity, and the spectral energy emission rate within the geometric-optics approximation.

This paper is organized as follows. In Sec.~\ref{sec:2} we present the BH spacetime and analyze the behaviour of the lapse function. The photon dynamics, including the photon sphere, shadow, effective potential, radial force, and the orbit equation, are studied in the subsections. The dynamics of massive neutral test particles, the derivation of the specific angular momentum and energy, and the stability conditions for circular orbits are discussed subsequently. Section~\ref{sec:thermo} is devoted to the thermodynamic analysis of the BH, and Sec.~\ref{sec:emission} addresses the spectral energy emission rate. We summarize our main findings in Sec.~\ref{sec:conclusions}. Throughout this paper, we adopt geometric units with $G = c = 1$ and, for numerical calculations, set the BH mass $M = 1$.

\section{Dyonic ModMax BH in KR-gravity with CS}\label{sec:2}

We consider a static and spherically symmetric line element that describes a dyonic
ModMax black hole in KR-gravity, with a cloud of strings acting as the matter source. The most general static, spherically symmetric metric compatible with these symmetries can be written as \cite{Ahmed2026a,Ahmed2026b},
\begin{equation}
    ds^2=-f(r)\,dt^2+\frac{dr^2}{f(r)}+r^2 (d\theta^2+\sin^2 \theta\, d\phi^2),
    \label{aa1}
\end{equation}
where the same function $f(r)$ appears in both the $tt$ and $rr$ metric components,
a feature characteristic of solutions that satisfy the condition $g_{tt}\,g_{rr}=-1$. The lapse function $f(r)$ is obtained by solving the Einstein field equations with three simultaneous sources, the ModMax electromagnetic field, the KR background tensor field, and the cloud of strings, and reads
\begin{equation}
f(r)=\frac{1-\alpha}{1-\ell}-\frac{2 M}{r}+ e^{-\gamma}\frac{(Q_e^2+Q_m^2)}{(1-\ell)^2 r^2}.\label{aa2}
\end{equation}
Here $\alpha\in(0,1)$ is the string-cloud parameter, $\ell$ is the dimensionless
KR-field parameter associated with the spontaneous breaking of local Lorentz symmetry, $M$ is the black hole mass, $Q_e$ is the electric charge, $Q_m$ is the magnetic charge, and $\gamma\geq 0$ is the ModMax nonlinearity parameter.

Each term in Eq.~\eqref{aa2} carries a distinct physical meaning. The first term, $(1-\alpha)/(1-\ell)$, is a constant that replaces the asymptotic unit value of the standard Schwarzschild metric: the cloud of strings reduces the asymptotic value through $\alpha$, while the KR field modulates it through $\ell$, both breaking the usual asymptotic flatness and inducing a global conical singularity. The second term is the standard Newtonian gravitational potential $-2M/r$. The third term represents the electromagnetic contribution, in which the factor $e^{-\gamma}$ is the ModMax screening of the combined dyonic charge squared, $Q_e^2+Q_m^2$: for $\gamma=0$ one recovers the standard RN charge term, whereas for $\gamma>0$ the effective charge is exponentially suppressed, and in the limit $\gamma\to\infty$ the metric approaches the Schwarzschild form. The additional factor $(1-\ell)^{-2}$ in the charge term reflects the rescaling of the electromagnetic sector by the KR field. Setting $\alpha=\ell =\gamma=0$ and $Q_m=0$ reproduces the standard Reissner-Nordstr\"{o}m metric.

The electromagnetic four-vector potential compatible with this geometry is given by
\begin{equation}
    A_{\mu}=\left(-\Phi(r),\,0,\,0,\,Q_m\,\cos \theta\right),\quad
    \Phi(r)=\frac{e^{-\gamma}\,Q_e}{r}.\label{aa3}
\end{equation}
The time component $A_t=-\Phi(r)$ is the electrostatic potential, which falls off as
$r^{-1}$ in the familiar Coulomb fashion but is screened by $e^{-\gamma}$ due to the
ModMax nonlinearity. The azimuthal component $A_\phi=Q_m\cos\theta$ is the standard
magnetic monopole potential, which generates a purely radial magnetic field
$B_r=Q_m/r^2$. The simultaneous presence of both $Q_e\neq 0$ and $Q_m\neq 0$
makes the configuration \emph{dyonic}, and it is precisely this duality between
electric and magnetic charges that are preserved by the SO(2) symmetry of the ModMax Lagrangian.

\begin{figure*}[ht]
    \centering
    \includegraphics[width=\textwidth]{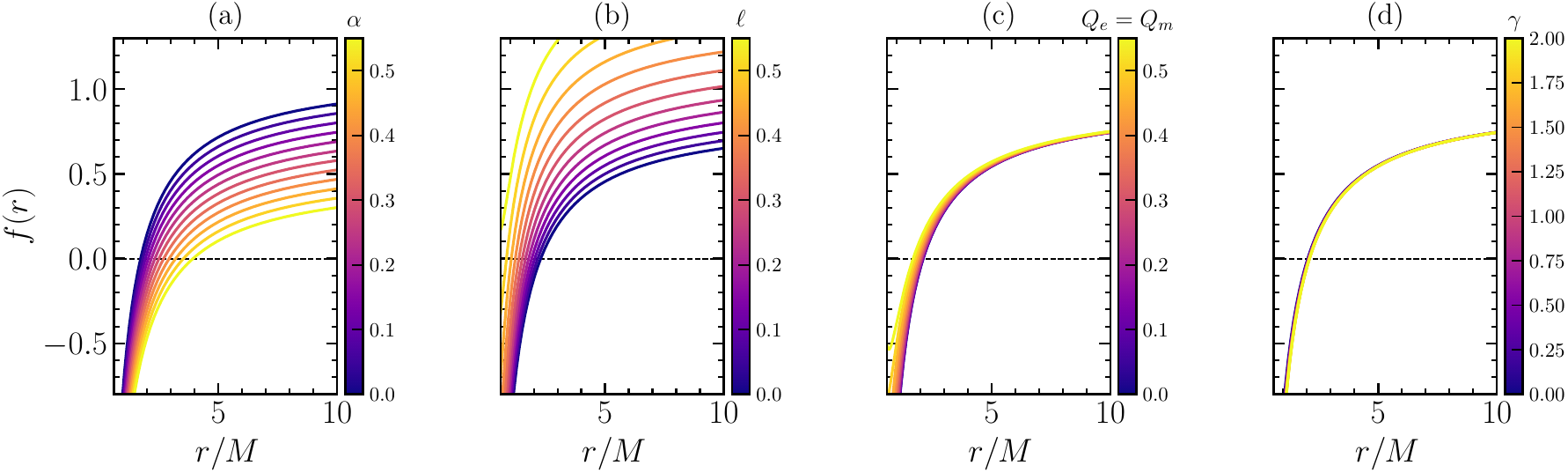}
    \caption{Lapse function $f(r)$ as a function of the radial coordinate $r/M$,
    for $M=1$. Each panel varies one parameter over 12 equally spaced values
    (color-coded from dark to light via a continuous colormap) while the remaining
    parameters are fixed at the reference set $\alpha=0.15$, $\ell=0.10$, $Q_e=0.30$, $Q_m=0.20$, $\gamma=0.10$. \textbf{(a)}~$\alpha \in [0.00,\,0.55]$ with $\ell=0.10$, $Q_e=0.30$, $Q_m=0.20$, $\gamma=0.10$. \textbf{(b)}~$\ell \in
    [0.00,\,0.55]$ with $\alpha=0.15$, $Q_e=0.30$, $Q_m=0.20$, $\gamma=0.10$.
    \textbf{(c)}~$Q_e=Q_m \in [0.00,\,0.55]$ with $\alpha=0.15$, $\ell=0.10$,
    $\gamma=0.10$. \textbf{(d)}~$\gamma \in [0.00,\,2.00]$ with $\alpha=0.15$,
    $\ell=0.10$, $Q_e=0.30$, $Q_m=0.20$. The dashed horizontal line marks $f(r)=0$.}
    \label{fig:metric}
\end{figure*}

The behavior of the lapse function $f(r)$ is displayed in Fig.~\ref{fig:metric}, where each panel varies one parameter at a time while the remaining ones are kept fixed at the reference values $\alpha=0.15$, $\ell=0.10$, $Q_e=0.30$, $Q_m=0.20$, and $\gamma=0.10$. The metric function exhibits the standard features of a Reissner-Nordstr\"{o}m-like black hole: a region of negative $f$ bounded by two horizons that merge into a single extremal horizon as the charges increase. Increasing the string cloud parameter $\alpha$ [panel~(a)] or the KR-field parameter $\ell$ [panel~(b)] effectively reduces the overall magnitude of the constant asymptotic value $f(r\to\infty)=(1-\alpha)/(1-\ell)$, which departs from unity due to the conical singularity induced by both fields, and shifts the horizon structure accordingly. Larger combined charge $Q_e=Q_m=Q$ [panel~(c)] pushes the two horizons closer together, eventually merging at the extremal
limit, while larger values of the ModMax nonlinearity parameter $\gamma$ [panel~(d)] reduce the effective electromagnetic contribution $e^{-\gamma}(Q_e^2+Q_m^2)$, rendering the metric progressively similar to the Schwarzschild case.

\section{Photon Dynamics}

To study the propagation of light rays in the spacetime~\eqref{aa1}, we work in the
equatorial plane $\theta=\pi/2$. This choice is not restrictive: since the background is spherically symmetric, any photon geodesic can be rotated into the equatorial plane without loss of generality. The motion is governed by the Lagrangian density 
\begin{equation}
    \mathcal{L}=\frac{1}{2}\left[-f(r)\,\dot t^2+\frac{\dot r^2}{f(r)}
    +r^2\, \dot \phi^2\right],\label{bb1}
\end{equation}
where an overdot denotes differentiation with respect to an affine parameter $\tau$
along the geodesic. For a null (photon) trajectory, the on-shell condition is
$2\mathcal{L}=0$, which will be used below to derive the radial equation of motion.

The metric~\eqref{aa1} is independent of the coordinates $t$ and $\phi$, so by the Euler-Lagrange equations, the corresponding conjugate momenta are constants of motion. These conserved quantities are the energy $\mathrm{E}$ and the angular momentum $\mathrm{L}$, defined respectively as
\begin{align}
    \mathrm{E}&=f(r)\,\dot t,\label{bb2}\\
    \mathrm{L}&=r^2\,\dot \phi.\label{bb3}
\end{align}
Here $\mathrm{E}$ is the energy of the photon as measured by a static observer at
spatial infinity, while $\mathrm{L}$ is the axial component of angular momentum.
Their ratio defines the \emph{impact parameter} $\beta\equiv\mathrm{L}/\mathrm{E}$,
which determines the asymptotic apparent position of the photon orbit.

Inverting Eqs.~\eqref{bb2} and~\eqref{bb3} for $\dot{t}$ and $\dot{\phi}$, and
substituting into the null condition $2\mathcal{L}=0$, we obtain the first-order
equations of motion for photon particles,
\begin{align}
    \dot t&=\mathrm{E}/f(r),\nonumber\\
    \dot \phi&=\mathrm{L}/r^2,\nonumber\\
    \dot r&=\sqrt{V_{\rm eff}},\label{bb4}
\end{align}
where the effective potential governing the radial motion is
\begin{equation}
    V_{\rm eff}=\mathrm{E}^2-\frac{\mathrm{L}^2}{r^2}\,f(r).\label{bb5}
\end{equation}
The sign of $V_{\rm eff}$ dictates whether radial motion is allowed
($V_{\rm eff}\geq 0$) or forbidden ($V_{\rm eff}<0$). Turning points of the
orbit correspond to $V_{\rm eff}=0$, while the maximum of $V_{\rm eff}(r)$ marks
the location of the \emph{unstable circular photon orbit}, also known as the photon
sphere, which plays a central role in determining the black hole shadow and the
gravitational lensing cross-section.

\begin{figure*}[ht]
\centering
\includegraphics[width=\textwidth]{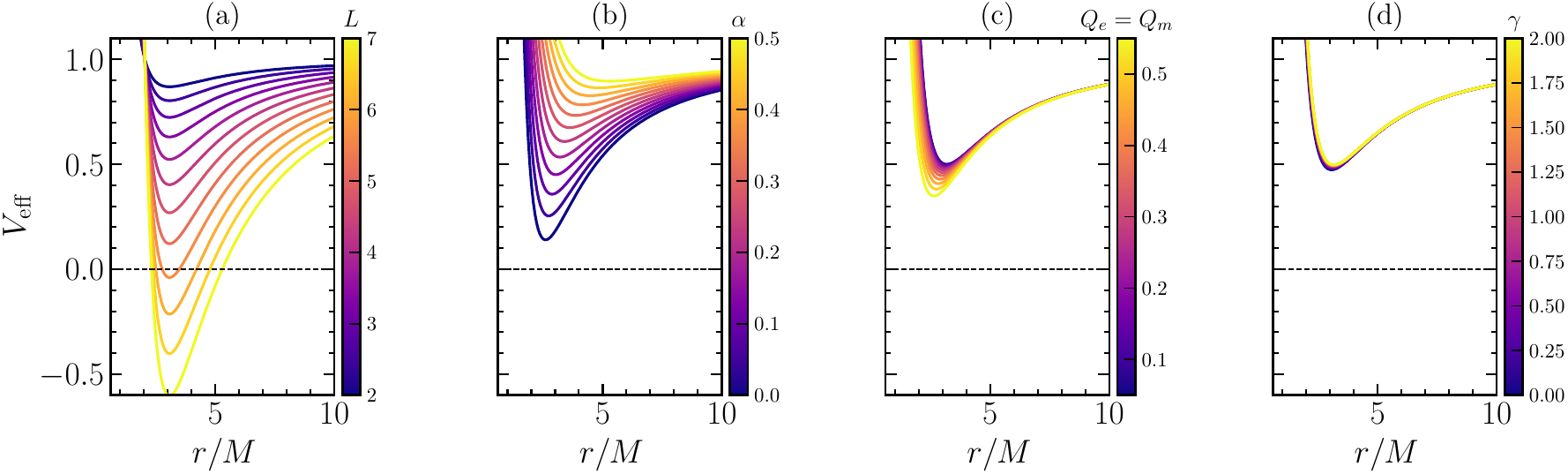}
\caption{Photon effective potential $V_{\rm eff}$ as a function of $r/M$ for
$M=1$ and $E=1$. Each panel shows 12 curves varying one parameter continuously
(colormap from dark to light). \textbf{(a)}~$L \in [2.0,\,7.0]\,M$ with $\alpha=0.15$, $\ell=0.10$, $Q_e=0.30$, $Q_m=0.20$, $\gamma=0.10$.
\textbf{(b)}~$\alpha \in [0.00,\,0.50]$ with $L=4\,M$, $\ell=0.10$,
$Q_e=0.30$, $Q_m=0.20$, $\gamma=0.10$. \textbf{(c)}~$Q_e=Q_m \in [0.05,\,0.55]$
with $L=4\,M$, $\alpha=0.15$, $\ell=0.10$, $\gamma=0.10$.
\textbf{(d)}~$\gamma \in [0.00,\,2.00]$ with $L=4\,M$, $\alpha=0.15$,
$\ell=0.10$, $Q_e=0.30$, $Q_m=0.20$. The dashed line marks $V_{\rm eff}=0$.}
\label{fig:veff}
\end{figure*}

The behaviour of the photon effective potential $V_{\rm eff}$ is illustrated in
Fig.~\ref{fig:veff}. Each panel varies one parameter over 12 equally spaced values
while fixing the remaining ones at $\alpha=0.15$, $\ell=0.10$, $Q_e=0.30$,
$Q_m=0.20$, $\gamma=0.10$ (and $L=4\,M$ except for panel~(a)). A characteristic
potential barrier is clearly visible in all panels, whose peak corresponds to the
unstable photon sphere. Increasing $L$ [panel~(a)] raises the potential barrier
uniformly, since $V_{\rm eff}$ scales as $L^2$. Increasing $\alpha$ [panel~(b)]
diminishes the asymptotic value of $V_{\rm eff}$ and shifts the barrier location
outward, reflecting the modified asymptotic flatness. Larger charges $Q$ [panel~(c)] push the barrier peak to smaller radii and slightly increase its height.
Finally, larger $\gamma$ [panel~(d)] lowers the barrier because the effective
electromagnetic contribution is suppressed by $e^{-\gamma}$, driving the spacetime
toward Schwarzschild geometry.

\subsection{Photon Sphere and Shadow}\label{subsec:shadow}

The photon sphere is defined as the locus of unstable circular photon orbits.
For a circular orbit at radius $r=r_s$ one requires simultaneously $\dot{r}=0$
(constant radius) and $\ddot{r}=0$ (no radial acceleration), i.e.\ the photon
sits at a turning point that is also an extremum of the effective potential.
Applying these two conditions to Eq.~\eqref{bb5} yields the single equivalent
requirement
\begin{equation}
    \frac{\partial}{\partial r}\!\left(\frac{f(r)}{r^2}\right)\bigg|_{r=r_s}=0.\label{bb6}
\end{equation}
Equation~\eqref{bb6} states that the ratio $f(r)/r^2$ is extremized at the photon
sphere, which follows directly from $V_{\rm eff}'=0$ once the conserved quantities
are factored out. Substituting the lapse function~\eqref{aa2} into Eq.~\eqref{bb6}
and simplifying, we obtain the algebraic condition
\begin{equation}
\frac{1-\alpha}{1-\ell}\,r_s^2-3 M r_s
+2e^{-\gamma}\frac{(Q_e^2+Q_m^2)}{(1-\ell)^2}=0.\label{bb7}
\end{equation}
This is a quadratic equation in $r_s$, whose physically relevant (larger) root gives the photon sphere radius
\begin{equation}
    r_s=\frac{1-\ell}{1-\alpha}\cdot\frac{1}{2}\left[3 M
    +\sqrt{9 M^2-8 e^{-\gamma}
    \frac{(1-\alpha)(Q_e^2+Q_m^2)}{(1-\ell)^3}}\right].\label{bb8}
\end{equation}
For $\alpha=\ell=\gamma=0$ and $Q_m=0$, Eq.~\eqref{bb8} reproduces the standard
RN result $r_s=(3M+\sqrt{9M^2-8Q_e^2})/2$, which in turn reduces to $r_s=3M$
for the Schwarzschild black hole when $Q_e\to 0$.

\begin{figure*}[ht]
    \centering
    \includegraphics[width=\textwidth]{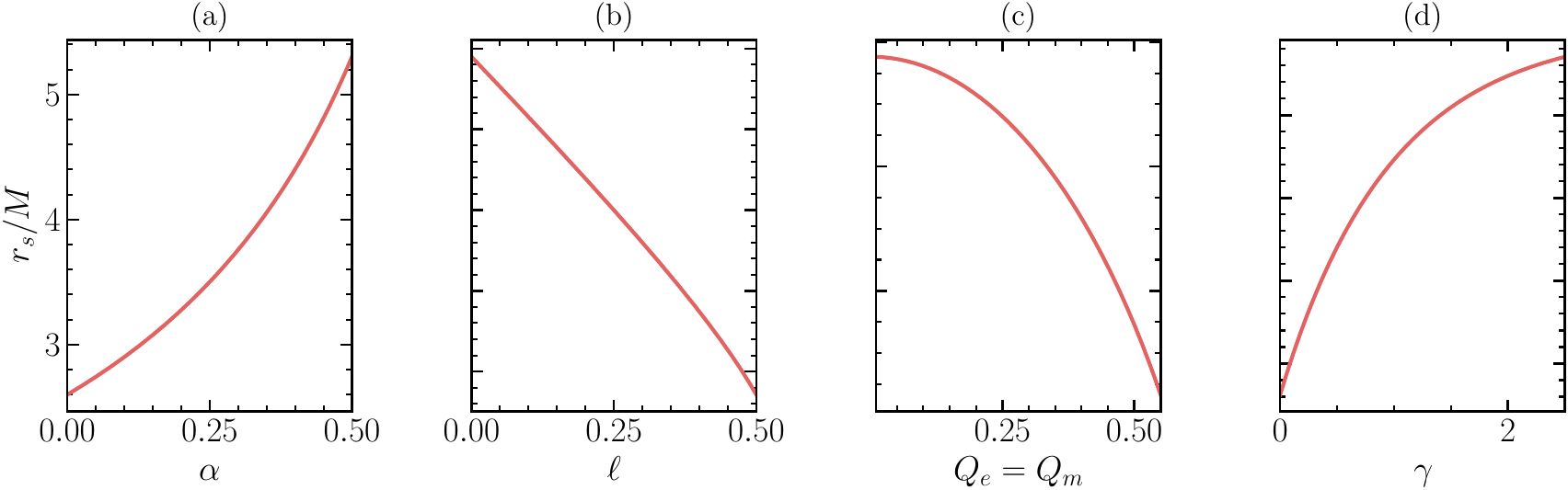}
    \caption{Photon sphere radius $r_s/M$ as a function of each free parameter,
    for $M=1$. The remaining parameters are fixed at $\alpha=0.15$, $\ell=0.10$,
    $Q_e=0.30$, $Q_m=0.20$, $\gamma=0.10$ in each panel.
    \textbf{(a)}~$r_s$ vs.\ $\alpha \in [0.00,\,0.50]$.
    \textbf{(b)}~$r_s$ vs.\ $\ell \in [0.00,\,0.50]$.
    \textbf{(c)}~$r_s$ vs.\ $Q_e=Q_m \in [0.01,\,0.55]$.
    \textbf{(d)}~$r_s$ vs.\ $\gamma \in [0.00,\,2.50]$.}
    \label{fig:rs}
\end{figure*}

The dependence of the photon sphere radius $r_s$ on the four free parameters is
shown in Fig.~\ref{fig:rs}. The photon sphere shrinks as the electric and magnetic
charges $Q$ grow [panel~(c)], consistent with the well-known Reissner-Nordstr\"{o}m
behaviour. The ModMax parameter $\gamma$ [panel~(d)] acts in the opposite direction:
since it suppresses the electromagnetic contribution via $e^{-\gamma}$, increasing
$\gamma$ gradually pushes $r_s$ toward the Schwarzschild value $r_s=3M$. Both
$\alpha$ [panel~(a)] and $\ell$ [panel~(b)] shift $r_s$ through their combined
prefactor $(1-\ell)/(1-\alpha)$: larger $\alpha$ increases the photon sphere radius, while larger $\ell$ decreases it.

Once the photon sphere radius is known, the critical impact parameter $\beta_c$
follows from the circular-orbit condition $V_{\rm eff}(r_s)=0$, which gives
$\mathrm{E}/\mathrm{L}=\sqrt{f(r_s)}/r_s$. Thus
\begin{align}
    \beta_c &= \frac{\mathrm{L}}{\mathrm{E}}=\frac{r_s}{\sqrt{f(r_s)}}\nonumber\\
    &=\frac{r^2_s}{\sqrt{\dfrac{1-\alpha}{1-\ell}\,r_s^2
    -2 M r_s+ e^{-\gamma}\dfrac{(Q_e^2+Q_m^2)}{(1-\ell)^2}}}.\label{bb9}
\end{align}
Photons with impact parameter $\beta<\beta_c$ are captured by the black hole,
while those with $\beta>\beta_c$ are deflected and escape to infinity; the marginal
orbit $\beta=\beta_c$ winds infinitely many times around the photon sphere. The photon sphere, and hence $\beta_c$, exists only when the discriminant in Eq.~\eqref{bb8} is non-negative, which imposes the constraint
\begin{equation}
M^2>\frac{8\,e^{-\gamma}(1-\alpha)(Q_e^2+Q_m^2)}{9\,(1-\ell)^3}.\label{bb_exist}
\end{equation}
This condition generalizes the familiar extremality bound of the RN black hole by
incorporating the effects of the string cloud, the KR field, and the ModMax
nonlinearity: the bound is relaxed (i.e.\ the photon sphere can exist for lighter
black holes) when $\gamma$ is large or when the combination $(1-\alpha)/(1-\ell)^3$
is small.

\begin{figure*}[ht]
    \centering
    \includegraphics[width=\textwidth]{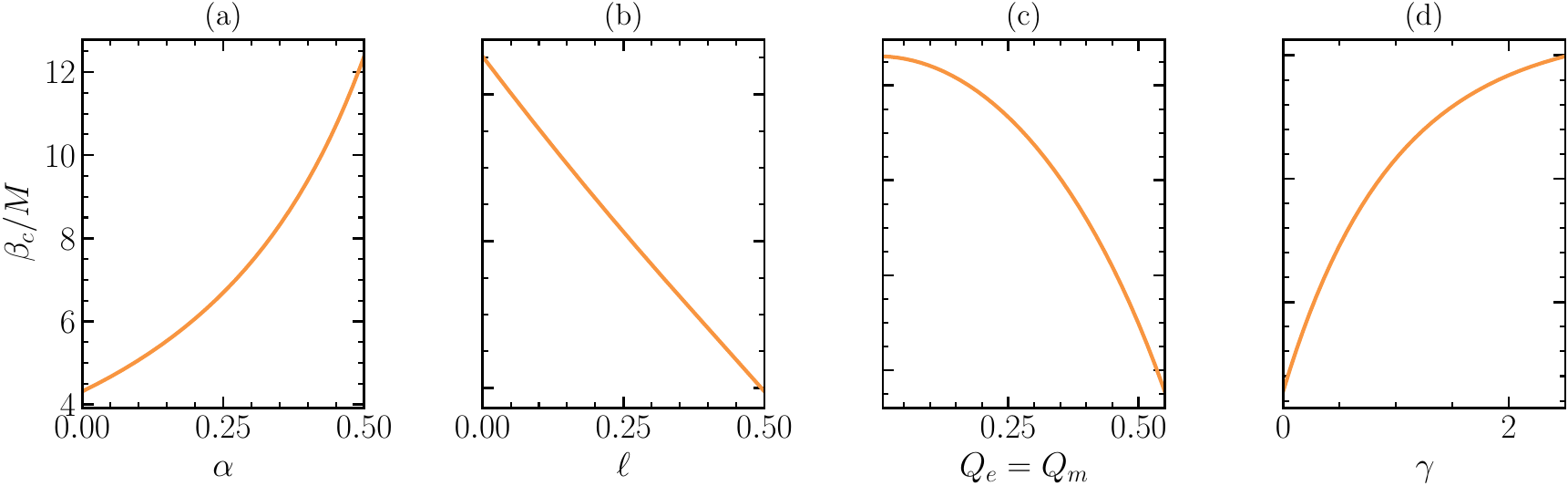}
    \caption{Critical impact parameter $\beta_c/M$ as a function of each free
    parameter, for $M=1$. The remaining parameters are fixed at $\alpha=0.15$,
    $\ell=0.10$, $Q_e=0.30$, $Q_m=0.20$, $\gamma=0.10$ in each panel.
    \textbf{(a)}~$\beta_c$ vs.\ $\alpha \in [0.00,\,0.50]$.
    \textbf{(b)}~$\beta_c$ vs.\ $\ell \in [0.00,\,0.50]$.
    \textbf{(c)}~$\beta_c$ vs.\ $Q_e=Q_m \in [0.01,\,0.55]$.
    \textbf{(d)}~$\beta_c$ vs.\ $\gamma \in [0.00,\,2.50]$.}
    \label{fig:betac}
\end{figure*}

The critical impact parameter $\beta_c$ as a function of the model parameters is
presented in Fig.~\ref{fig:betac}. The qualitative trends closely follow those of
$r_s$, since $\beta_c \propto r_s^2/\sqrt{f(r_s)}$. Larger charges [panel~(c)]
reduce $\beta_c$, leading to smaller effective shadow sizes. The ModMax parameter
$\gamma$ [panel~(d)] monotonically increases $\beta_c$ as the electromagnetic
coupling is suppressed, approaching the Schwarzschild critical impact parameter
$\beta_c = 3\sqrt{3}\,M$. The string cloud and KR parameters modulate $\beta_c$
through the asymptotic prefactor and the location of the photon sphere, as seen in
panels~(a) and~(b), respectively.

Before computing the shadow radius, we note the asymptotic behavior of the metric
function. As $r\to\infty$, the lapse function approaches the constant value
\begin{equation}
    f(r)\to\frac{1-\alpha}{1-\ell}\equiv f_\infty,
\end{equation}
which differs from unity whenever $\alpha\neq\ell$. This departure signals the
absence of asymptotic flatness: both the cloud of strings and the KR field modify
the global geometry and induce a conical singularity at spatial infinity. As a
consequence, the shadow formula must account for the non-flat asymptotics, as we
now show.

For a static observer located at radial position $r_O$, the angular radius of the black hole shadow is determined by the ratio of the metric function evaluated at the photon sphere and at the observer. Using the standard result of Ref.~\cite{Volker2022}, the angular size of the shadow satisfies
\begin{equation}
    \sin^2\theta_{\rm sh}=\frac{h^2(r_s)}{h^2(r_O)},\quad
    h^2(r)\equiv\frac{r^2}{f(r)},\label{bb10}
\end{equation}
where $h(r)$ is the impact-parameter function and the ratio $h(r_s)/h(r_O)$
compares the light-bending at the photon sphere to the geometry at the observer.
In the small-angle limit, the observed shadow radius is
\begin{equation}
    R_{\rm sh} \simeq r_O\,\theta_{\rm sh}.\label{bb11}
\end{equation}
Substituting the metric function~\eqref{aa2} into Eqs.~\eqref{bb10}--\eqref{bb11}
yields the exact expression valid for any finite observer distance $r_O$,
\begin{equation}
R_{\rm sh}=r_s\,\sqrt{\frac{\dfrac{1-\alpha}{1-\ell}-\dfrac{2M}{r_O}
+ e^{-\gamma}\dfrac{(Q_e^2+Q_m^2)}{(1-\ell)^2 r_O^2}}
{\dfrac{1-\alpha}{1-\ell}-\dfrac{2M}{r_s}
+ e^{-\gamma}\dfrac{(Q_e^2+Q_m^2)}{(1-\ell)^2 r_s^2}}}.\label{bb12}
\end{equation}
The denominator of the radicand is simply $f(r_s)$, while the numerator is $f(r_O)$; the square-root factor therefore equals $\sqrt{f(r_O)/f(r_s)}$, confirming the geometric origin of the formula. In the distant-observer limit $r_O\to\infty$, the numerator approaches $f_\infty=(1-\alpha)/(1-\ell)$ and Eq.~\eqref{bb12} simplifies to the compact expression
\begin{equation}
R_{\rm sh}=\sqrt{\frac{1-\alpha}{1-\ell}}\,\beta_c,\label{bb13}
\end{equation}
where $\beta_c$ is given by Eq.~\eqref{bb9}. The factor $\sqrt{f_\infty}$,
absent in the standard asymptotically flat case, encodes the joint rescaling of the
apparent shadow size by both the string cloud and the KR field.

\begin{figure*}[ht]
    \centering
    \includegraphics[width=\textwidth]{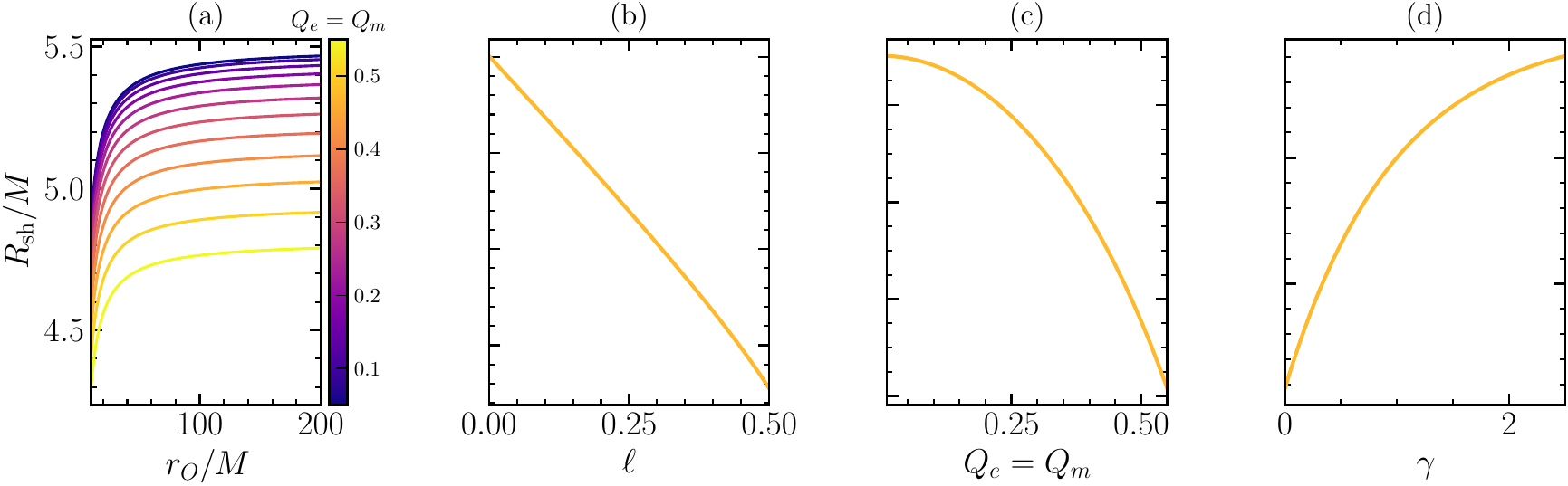}
    \caption{Black hole shadow radius $R_{\rm sh}/M$ for $M=1$.
    \textbf{(a)}~$R_{\rm sh}$ as a function of the observer distance
    $r_O\in[10,\,200]\,M$ for 12 values of $Q_e=Q_m\in[0.05,\,0.55]$ (colormap),
    with $\alpha=0.15$, $\ell=0.10$, $\gamma=0.10$.
    \textbf{(b)}~$R_{\rm sh}$ for a distant observer vs.\
    $\ell\in[0.00,\,0.50]$, with $\alpha=0.15$, $Q_e=0.30$, $Q_m=0.20$,
    $\gamma=0.10$.
    \textbf{(c)}~$R_{\rm sh}$ vs.\ $Q_e=Q_m\in[0.01,\,0.55]$, with $\alpha=0.15$,
    $\ell=0.10$, $\gamma=0.10$.
    \textbf{(d)}~$R_{\rm sh}$ vs.\ $\gamma\in[0.00,\,2.50]$, with $\alpha=0.15$,
    $\ell=0.10$, $Q_e=0.30$, $Q_m=0.20$.}
    \label{fig:shadow}
\end{figure*}

Figure~\ref{fig:shadow} displays the shadow radius $R_{\rm sh}$ under different
settings. Panel~(a) shows $R_{\rm sh}$ as a function of the observer distance $r_O$
for 12 equally spaced values of the combined charge $Q_e=Q_m=Q\in[0.05,\,0.55]$,
with $\alpha=0.15$, $\ell=0.10$, and $\gamma=0.10$. As expected, $R_{\rm sh}$
converges to the asymptotic distant-observer value for large $r_O$, and higher charges produce systematically smaller shadows. Panels~(b), (c), and~(d) display
$R_{\rm sh}$ for a distant observer as a function of $\ell$, $Q$, and $\gamma$, respectively, confirming the monotonic trends already established for $\beta_c$: the shadow shrinks with increasing charge or decreasing ModMax coupling, while the
KR-field parameter $\ell$ and the asymptotic geometry govern the overall scale
through the prefactor $\sqrt{(1-\alpha)/(1-\ell)}$.

\subsection{Photon Trajectory and Effective Force}

The shape of a photon orbit is described by the trajectory equation, which relates the radial and angular coordinates. Dividing $\dot{r}$ by $\dot{\phi}$ from
Eqs.~\eqref{bb4} and using $\beta=\mathrm{L}/\mathrm{E}$, we obtain the orbit equation in the form
\begin{align}
\left(\frac{dr}{d\phi}\right)^2&=r^4\left[\frac{1}{\beta^2}-\frac{f(r)}{r^2}\right]\notag\\&=r^4\!\left[\frac{1}{\beta^2}-\frac{1}{r^2}\!\left(\frac{1-\alpha}{1-\ell} -\frac{2 M}{r}+e^{-\gamma}\frac{Q_e^2+Q_m^2}{(1-\ell)^2 r^2}\right)\right].\label{cc1}
\end{align}

To cast this into a standard form amenable to perturbative treatment, we introduce the inverse-radial variable $u(\phi)=1/r(\phi)$, so that $dr/d\phi = -(1/u^2)\,du/d\phi$. Substituting and rearranging, Eq.~\eqref{cc1} becomes
\begin{equation}
\left(\frac{du}{d\phi}\right)^2+\frac{1-\alpha}{1-\ell}\,u^2=\frac{1}{\beta^2}+2 M u^3-e^{-\gamma}\frac{(Q_e^2+Q_m^2)}{(1-\ell)^2}\,u^4.\label{cc2}
\end{equation}
Differentiating both sides w. r. to $\phi$ and after simplification results
\begin{equation}
\frac{d^2u}{d\phi^2}+\frac{1-\alpha}{1-\ell}\,u=3 M u^2-2\,e^{-\gamma}\frac{(Q_e^2+Q_m^2)}{(1-\ell)^2}\,u^3.\label{cc2a}
\end{equation}

The left-hand side of Eq.~\eqref{cc2} resembles a one-dimensional harmonic oscillator in $u$ with a modified frequency $\sqrt{(1-\alpha)/(1-\ell)}$, while the right-hand side contains the source terms: a constant $1/\beta^2$ (analogous to a driving term), the relativistic cubic $2Mu^3$ familiar from the Schwarzschild
case, and the electromagnetic quartic $-e^{-\gamma}(Q_e^2+Q_m^2)u^4/(1-\ell)^2$ that is a direct signature of the dyonic ModMax field. Differentiating
Eq.~\eqref{cc2} once with respect to $\phi$ yields the Binet equation, which can be solved perturbatively to obtain the deflection angle and orbital precession.

\begin{figure*}[ht]
    \centering
    \includegraphics[width=\textwidth]{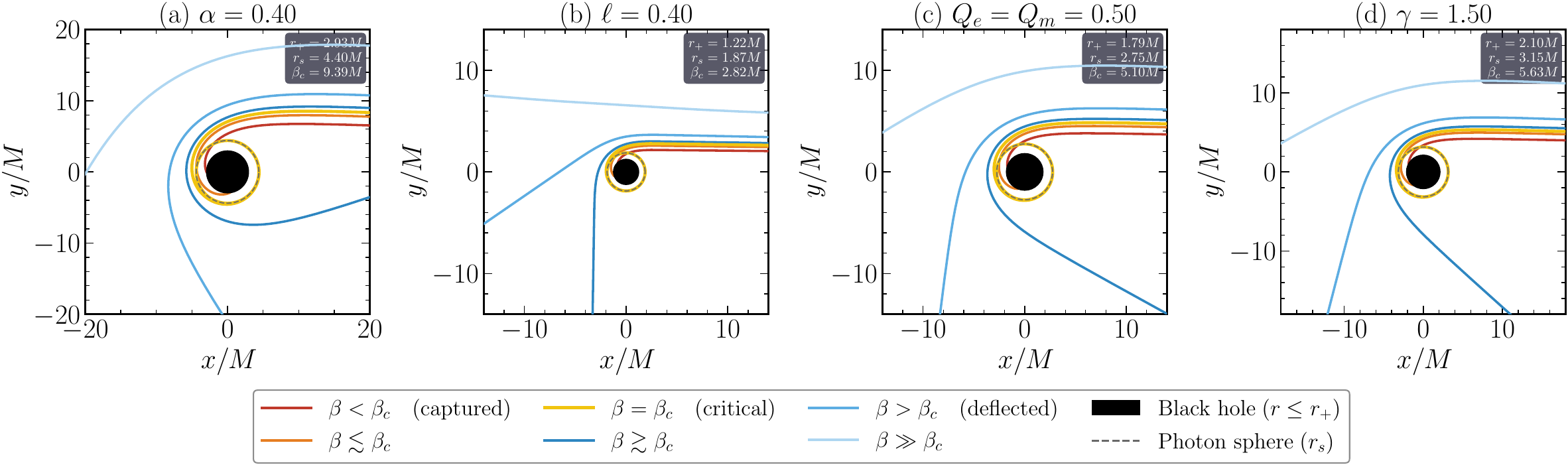}
    \caption{Photon trajectories in the equatorial plane, obtained by numerically
    integrating the Binet equation~\eqref{cc2}, for $M=1$.
    Each panel fixes three parameters at the reference set
    $\alpha=0.15$, $\ell=0.10$, $Q_e=0.30$, $Q_m=0.20$, $\gamma=0.10$
    and assigns one parameter a representative value:
    \textbf{(a)}~$\alpha=0.40$,
    \textbf{(b)}~$\ell=0.40$,
    \textbf{(c)}~$Q_e=Q_m=0.50$,
    \textbf{(d)}~$\gamma=1.50$.
    Red/orange curves: captured photons ($\beta<\beta_c$).
    Gold curve: critical orbit ($\beta=\beta_c$), winding asymptotically
    around the photon sphere. Blue curves: deflected photons ($\beta>\beta_c$).
    The filled black circle denotes the black hole interior ($r\leq r_+$) and
    the dashed gray circle marks the photon sphere at $r=r_s$.
    The values of $r_+$, $r_s$, and $\beta_c$ for each panel are given in the
    inset boxes.}
    \label{fig:orbits}
\end{figure*}

The photon orbits obtained by numerically integrating the equation~\eqref{cc2}---the second-order form of Eq.~\eqref{cc3}-are displayed in Fig.~\ref{fig:orbits}. Each panel fixes three parameters at the reference values and assigns a representative value to the fourth, allowing the
structural changes in the orbit topology to be read off directly. In every panel, trajectories are shown for six values of the impact parameter $\beta$: two with $\beta<\beta_c$ (red and orange, captured orbits), one precisely at $\beta=\beta_c$ (gold, critical orbit), and three with $\beta>\beta_c$ (blue shades, deflected orbits). The black filled circle represents the black hole interior ($r\leq r_+$) and the dashed gray circle marks the photon sphere at $r=r_s$.

The critical orbit ($\beta=\beta_c$) asymptotically winds around the photon sphere,
which is the characteristic signature of the unstable circular photon orbit: in the
absence of perturbations, a photon with exactly $\beta=\beta_c$ spirals infinitely
many times at $r=r_s$. Orbits with $\beta\lesssim\beta_c$ (orange) pass very close
to the photon sphere before being captured, producing the longest trajectories among
the captured set. Conversely, orbits with $\beta\gtrsim\beta_c$ undergo the largest
deflection angles among the escaping set, demonstrating the divergence of the
scattering angle as $\beta\to\beta_c^+$.

Panel~(a) ($\alpha=0.40$) shows the effect of a strong string cloud. Since $\alpha$
enters $r_s$ and $\beta_c$ through the prefactor $\lambda=(1-\alpha)/(1-\ell)$,
larger $\alpha$ enlarges both the photon sphere and the critical impact parameter
substantially---visible as an expanded dark disk and a wider capture cross section
compared to the reference. Panel~(b) ($\ell=0.40$) illustrates the complementary
effect of the KR field: increasing $\ell$ decreases $\lambda$ and therefore
increases $r_s$ as well, while simultaneously modifying the asymptotic geometry
so that $\beta_c$ is also enlarged; the orbit structure is qualitatively similar
to panel~(a) but with a smaller conical deficit. Panel~(c) ($Q_e=Q_m=0.50$) shows
the effect of large dyonic charges: the enhanced electromagnetic repulsion
(screened by $e^{-\gamma}$) pushes $r_s$ inward, reducing both the photon sphere
radius and $\beta_c$, so the capture cross section shrinks and deflected orbits
pass closer to the black hole before escaping. Panel~(d) ($\gamma=1.50$) demonstrates
strong ModMax screening: the exponential suppression of the effective charge
$e^{-\gamma}(Q_e^2+Q_m^2)$ drives $r_s$ toward the Schwarzschild value $3M$ and
$\beta_c$ toward $3\sqrt{3}M$, so the orbit structure approaches that of the
uncharged Schwarzschild case; the photon sphere is larger than in panel~(c), and
the capture region is correspondingly wider.

The effective radial force on a photon is defined as the gradient of the effective
potential,
\begin{equation}
    \mathcal{F}=-\frac{1}{2}\,\frac{\partial}{\partial r}
    \!\left(\frac{\mathrm{L}^2}{r^2}\,f(r)\right).\label{cc3}
\end{equation}
The factor $\mathrm{L}^2 f(r)/r^2 = \mathrm{L}^2/h^2(r)$ is proportional to the
angular part of $V_{\rm eff}$, so $\mathcal{F}$ measures the net radial push or pull experienced by the photon due to the combined gravitational and electromagnetic fields. A positive $\mathcal{F}$ corresponds to a centrifugal (outward) push, while a negative $\mathcal{F}$ corresponds to a gravitational (inward) pull. Substituting the metric function~\eqref{aa2} and carrying out the derivative, we find
\begin{equation}
\mathcal{F}=\frac{\mathrm{L}^2}{r^3}\left[\frac{1-\alpha}{1-\ell}
-\frac{3 M}{r}
+\frac{2\,e^{-\gamma}(Q_e^2+Q_m^2)}{(1-\ell)^2 r^2}\right].\label{cc4}
\end{equation}
Setting $\mathcal{F}=0$ in Eq.~\eqref{cc4} reproduces exactly the photon sphere
condition~\eqref{bb7}, confirming that the photon sphere is the equilibrium point where the centrifugal and gravitational effects balance. The three terms inside the bracket have a clear structure: the first, $(1-\alpha)/(1-\ell)$, is the asymptotic centrifugal term enhanced (or reduced) by the conical geometry; the second, $-3M/r$, is the gravitational attraction term (which in the Schwarzschild
case alone would give $r_s=3M$); and the third is the electromagnetic repulsion,
screened by $e^{-\gamma}$ and rescaled by $(1-\ell)^{-2}$.

\begin{figure*}[ht]
    \centering
    \includegraphics[width=\textwidth]{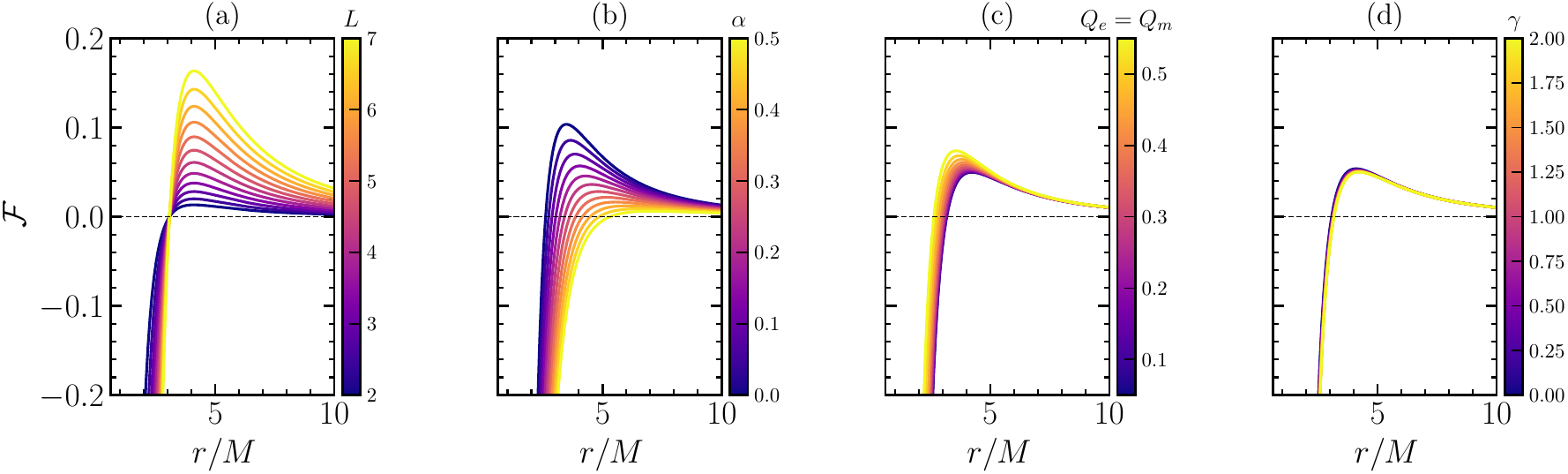}
    \caption{Effective radial force $\mathcal{F}$ on photons as a function of
    $r/M$, for $M=1$. Each panel shows 12 curves with one parameter varying
    continuously (colormap).
    \textbf{(a)}~$L\in[2.0,\,7.0]\,M$ with $\alpha=0.15$, $\ell=0.10$,
    $Q_e=0.30$, $Q_m=0.20$, $\gamma=0.10$.
    \textbf{(b)}~$\alpha\in[0.00,\,0.50]$ with $L=4\,M$, $\ell=0.10$,
    $Q_e=0.30$, $Q_m=0.20$, $\gamma=0.10$.
    \textbf{(c)}~$Q_e=Q_m\in[0.05,\,0.55]$ with $L=4\,M$, $\alpha=0.15$,
    $\ell=0.10$, $\gamma=0.10$.
    \textbf{(d)}~$\gamma\in[0.00,\,2.00]$ with $L=4\,M$, $\alpha=0.15$,
    $\ell=0.10$, $Q_e=0.30$, $Q_m=0.20$.
    The dashed line marks $\mathcal{F}=0$.}
    \label{fig:force}
\end{figure*}

The effective radial force $\mathcal{F}$ acting on photons is shown in
Fig.~\ref{fig:force}. The zero of $\mathcal{F}$ marks the location of the photon
sphere, where the centrifugal and gravitational effects balance. Inside this radius
the force is attractive ($\mathcal{F}<0$), while outside it is repulsive
($\mathcal{F}>0$), which is consistent with the instability of circular photon
orbits. Panel~(a) shows that larger angular momentum $L$ amplifies the magnitude of
the force without shifting the equilibrium radius, as expected from the overall
$\mathrm{L}^2/r^3$ prefactor in Eq.~\eqref{cc4}. Increasing the string cloud
parameter $\alpha$ [panel~(b)] shifts the zero of $\mathcal{F}$ outward, in
agreement with the enlarged photon sphere. Growing charges $Q$ [panel~(c)] move the
equilibrium point inward, while larger $\gamma$ [panel~(d)] weakens the
electromagnetic repulsion term, again shifting the force zero toward $r_s = 3M$.

\section{Dynamics of Neutral Particles}\label{sec:neutral}

The analysis of massive test-particle orbits provides complementary information to the photon-sphere results above, in particular, the location of the innermost stable circular orbit (ISCO), which governs the structure of accretion discs and determines the radiative efficiency of the black hole. For a massive, neutral test particle of rest mass $m$ moving in the curved spacetime~\eqref{aa1}, the geodesic motion follows from the Lagrangian density
\begin{equation}
    \mathbb{L}=\frac{1}{2}\,m\,g_{\mu\nu}\frac{dx^{\mu}}{d\lambda}\,
    \frac{dx^{\nu}}{d\lambda},\label{dd1}
\end{equation}
where $\lambda$ is the proper time along the worldline. Unlike the photon case, the on-shell condition is now $2\mathbb{L}/m = g_{\mu\nu}\dot{x}^\mu\dot{x}^\nu = -1$ (timelike geodesic), which introduces a mass term into the effective potential.

Proceeding exactly as in Sec.~\ref{sec:neutral}, the symmetries of the metric yield
the conserved specific energy $\mathcal{E}$ and specific angular momentum $\mathcal{L}$ (both per unit mass $m$), and the equations of motion reduce to the first-order system
\begin{align}
    \frac{dt}{d\lambda}&=\mathcal{E}/f(r),\nonumber\\
    \frac{d\phi}{d\lambda}&=\mathcal{L}/r^2,\nonumber\\
    \frac{dr}{d\lambda}&=\sqrt{U_{\rm eff}},\label{dd2}
\end{align}
where the effective potential is
\begin{equation}
    U_{\rm eff}=\mathcal{E}^2-\left(1+\frac{\mathcal{L}^2}{r^2}\right)f(r).\label{dd3}
\end{equation}
Compared to the photon potential $V_{\rm eff}$ in Eq.~\eqref{bb5}, the extra factor of unity inside the parentheses is the rest-mass contribution. This term lowers the effective potential at large radii, producing a local minimum at finite $r$ that is absent for photons and that signals the existence of bound, stable circular orbits.

\begin{figure*}[ht]
    \centering
    \includegraphics[width=\textwidth]{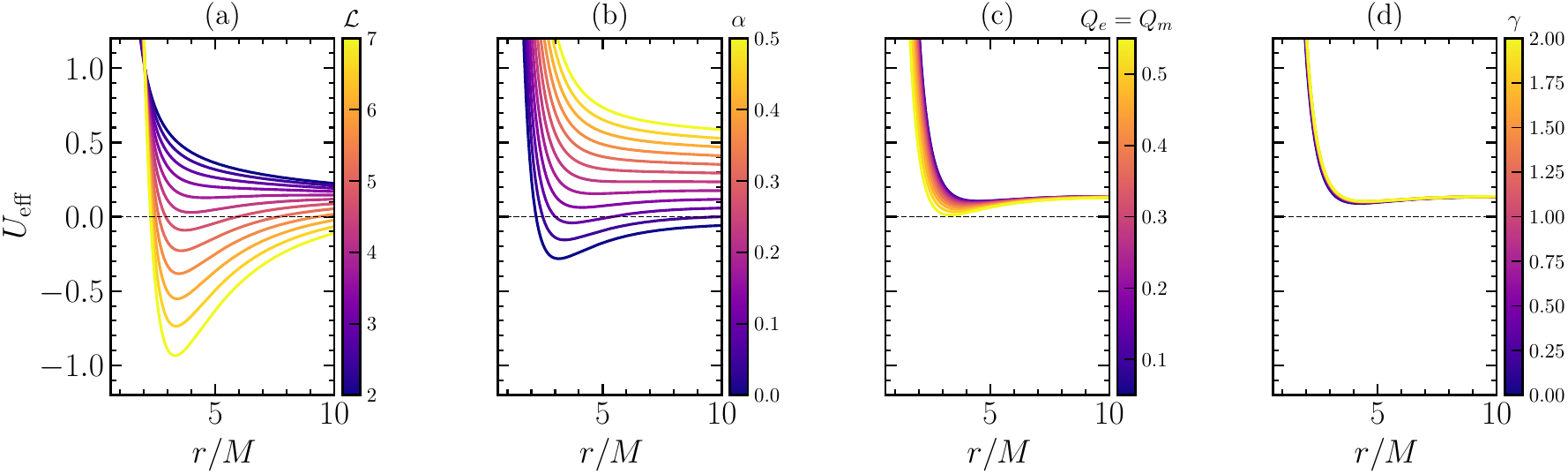}
    \caption{Effective potential $U_{\rm eff}$ for massive neutral test particles as
    a function of $r/M$, for $M=1$ and $\mathcal{E}=1$. Each panel shows 12 curves
    varying one parameter continuously (colormap).
    \textbf{(a)}~$\mathcal{L}\in[2.0,\,7.0]\,M$ with $\alpha=0.15$, $\ell=0.10$,
    $Q_e=0.30$, $Q_m=0.20$, $\gamma=0.10$.
    \textbf{(b)}~$\alpha\in[0.00,\,0.50]$ with $\mathcal{L}=4\,M$, $\ell=0.10$,
    $Q_e=0.30$, $Q_m=0.20$, $\gamma=0.10$.
    \textbf{(c)}~$Q_e=Q_m\in[0.05,\,0.55]$ with $\mathcal{L}=4\,M$, $\alpha=0.15$,
    $\ell=0.10$, $\gamma=0.10$.
    \textbf{(d)}~$\gamma\in[0.00,\,2.00]$ with $\mathcal{L}=4\,M$, $\alpha=0.15$,
    $\ell=0.10$, $Q_e=0.30$, $Q_m=0.20$.
    The dashed line marks $U_{\rm eff}=0$.}
    \label{fig:ueff}
\end{figure*}

Figure~\ref{fig:ueff} presents the effective potential $U_{\rm eff}$ for massive
neutral particles. In contrast to the photon potential, $U_{\rm eff}$ for massive
particles exhibits a local minimum at a finite radius, signalling the existence of
stable circular orbits (ISCO). The potential barrier and the local minimum are both
sensitive to $\mathcal{L}$: larger angular momentum raises the centrifugal barrier
and deepens the potential well [panel~(a)]. Increasing $\alpha$ [panel~(b)] shifts
both features outward and increases the depth of the well, indicating that the ISCO
migrates to larger radii in the presence of a stronger cloud of strings. Larger
combined charges $Q$ [panel~(c)] raise the local maximum of the barrier at
intermediate radii due to the enhanced electromagnetic repulsion. The ModMax
parameter $\gamma$ [panel~(d)] modulates the strength of the electromagnetic
contribution: as $\gamma$ grows the potential approaches the Schwarzschild
massive-particle case.

For a circular orbit at radius $r$, the particle must satisfy $dr/d\lambda=0$ (constant radius) and $d^2r/d\lambda^2=0$ (no radial acceleration), which translates into the simultaneous conditions
\begin{equation}
    \mathcal{E}^2=\left(1+\frac{\mathcal{L}^2}{r^2}\right)f(r),\label{dd4}
\end{equation}
\begin{equation}
    \frac{\partial U_{\rm eff}}{\partial r}=0.\label{dd5}
\end{equation}
Equation~\eqref{dd4} is the energy constraint (the orbit lies on the $U_{\rm eff}=0$ surface), while Eq.~\eqref{dd5} states that the orbit is at an extremum of the potential. Solving these two conditions simultaneously for the specific angular
momentum and specific energy of a circular orbit (hereafter "specific" quantities
for circular orbits are denoted with the subscript "sp"), we find
\begin{equation}
    \mathcal{L}_{\rm sp}=r\,\sqrt{\frac{r\,f'(r)}{2\,f(r)-r\,f'(r)}},\label{dd5a}
\end{equation}
\begin{equation}
    \mathcal{E}_{\rm sp}=\sqrt{\frac{2\,f(r)^2}{2\,f(r)-r\,f'(r)}}.\label{dd6}
\end{equation}

\begin{figure*}[ht]
    \centering
    \includegraphics[width=\textwidth]{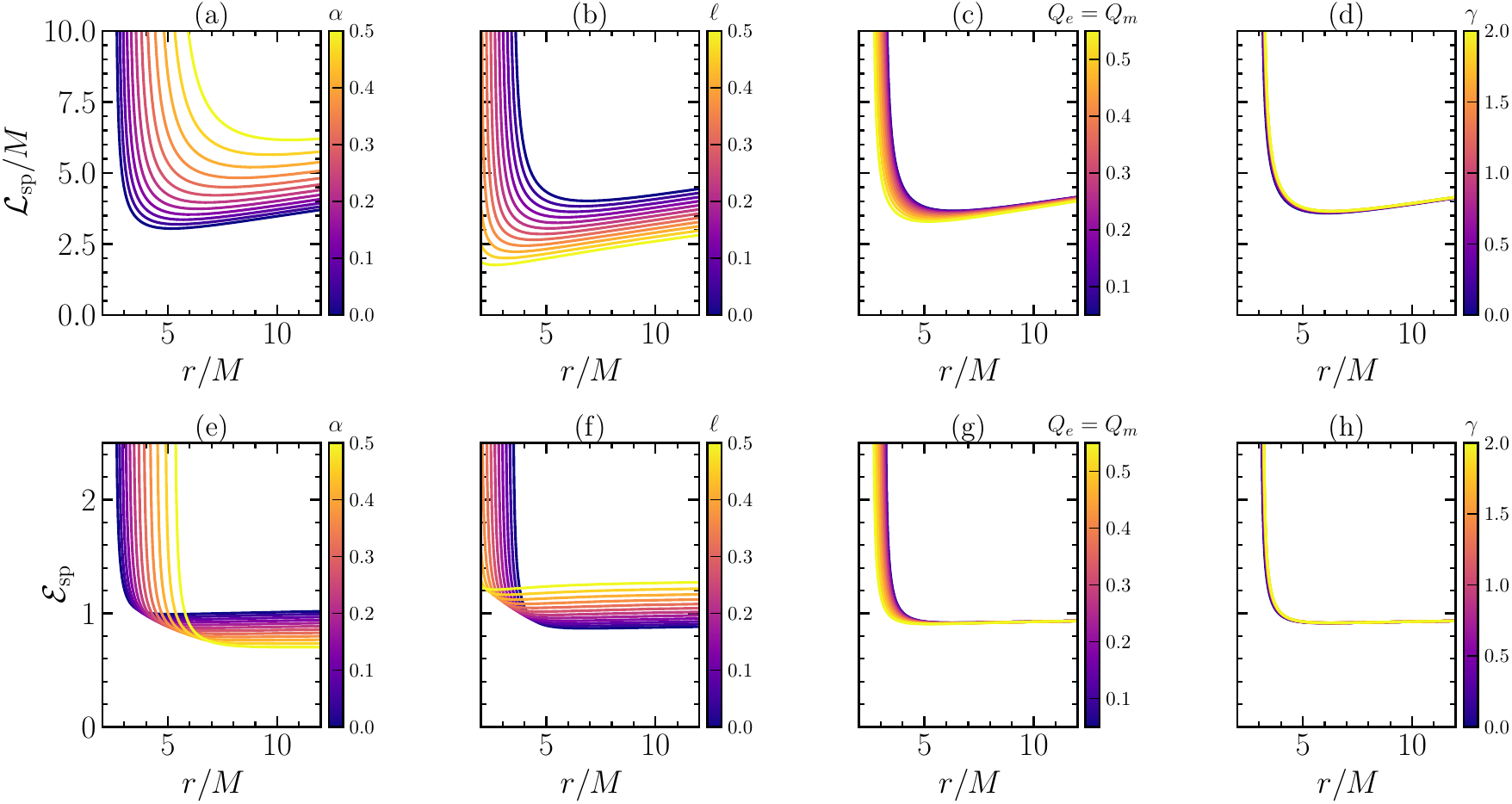}
    \caption{Specific angular momentum $\mathcal{L}_{\rm sp}/M$ (upper row,
    panels~(a)--(d)) and specific energy $\mathcal{E}_{\rm sp}$ (lower row,
    panels~(e)--(h)) for circular orbits of massive test particles as a function
    of $r/M$, for $M=1$. Each column varies one parameter over 12 equally spaced
    values (colormap), while the remaining parameters are fixed at $\alpha=0.15$,
    $\ell=0.10$, $Q_e=0.30$, $Q_m=0.20$, $\gamma=0.10$.
    \textbf{(a),(e)}~$\alpha\in[0.00,\,0.50]$.
    \textbf{(b),(f)}~$\ell\in[0.00,\,0.50]$.
    \textbf{(c),(g)}~$Q_e=Q_m\in[0.05,\,0.55]$.
    \textbf{(d),(h)}~$\gamma\in[0.00,\,2.00]$.}
    \label{fig:LspEsp}
\end{figure*}

\begin{table*}[t]
\centering
\caption{ISCO radius $r_\mathrm{ISCO}/M$ for different values of $q(=Q_e=Q_m)$, $\ell$, and $\alpha$.}
\footnotesize

\begin{minipage}[t]{0.48\textwidth}
\centering
\begin{tabular}{c c c c c c c}
\hline
\multicolumn{7}{c}{$\gamma=0.2$} \\
\hline
$\ell$ & $\alpha$ & $q/M=0.1$ & $q/M=0.2$ & $q/M=0.3$ & $q/M=0.4$ & $q/M=0.5$ \\
\hline
\multirow{4}{*}{$-0.2$}
& 0.05 & 7.56186 & 7.51028 & 7.42317 & 7.29870 & 7.13395 \\
& 0.10 & 7.98292 & 7.93136 & 7.84435 & 7.72016 & 7.55607 \\
& 0.15 & 8.45351 & 8.40197 & 8.31506 & 8.19115 & 8.02772 \\
& 0.20 & 8.98292 & 8.93140 & 8.84459 & 8.72097 & 8.55818 \\
\hline
\multirow{4}{*}{$-0.1$}
& 0.05 & 6.92703 & 6.86549 & 6.76115 & 6.61103 & 6.41027 \\
& 0.10 & 7.31299 & 7.25149 & 7.14731 & 6.99765 & 6.79800 \\
& 0.15 & 7.74437 & 7.68290 & 7.57887 & 7.42967 & 7.23111 \\
& 0.20 & 8.22967 & 8.16823 & 8.06436 & 7.91561 & 7.71812 \\
\hline
\multirow{4}{*}{$0.1$}
& 0.05 & 5.65377 & 5.56103 & 5.40131 & 5.16514 & 4.83426 \\
& 0.10 & 5.96957 & 5.87692 & 5.71766 & 5.48300 & 5.15630 \\
& 0.15 & 6.32252 & 6.22997 & 6.07117 & 5.83797 & 5.51528 \\
& 0.20 & 6.71958 & 6.62713 & 6.46879 & 6.23702 & 5.91818 \\
\hline
\end{tabular}
\end{minipage}
\hfill
\begin{minipage}[t]{0.48\textwidth}
\centering
\begin{tabular}{c c c c c c c}
\hline
\multicolumn{7}{c}{$\gamma=0.4$} \\
\hline
$\ell$ & $\alpha$ & $q/M=0.1$ & $q/M=0.2$ & $q/M=0.3$ & $q/M=0.4$ & $q/M=0.5$ \\
\hline
\multirow{4}{*}{$-0.2$}
& 0.05 & 7.56496 & 7.52279 & 7.45176 & 7.35064 & 7.21760 \\
& 0.10 & 7.98602 & 7.94386 & 7.87289 & 7.77196 & 7.63934 \\
& 0.15 & 8.45661 & 8.41447 & 8.34356 & 8.24282 & 8.11062 \\
& 0.20 & 8.98602 & 8.94389 & 8.87305 & 8.77250 & 8.64071 \\
\hline
\multirow{4}{*}{$-0.1$}
& 0.05 & 6.93072 & 6.88044 & 6.79545 & 6.67383 & 6.51252 \\
& 0.10 & 7.31669 & 7.26643 & 7.18154 & 7.06022 & 6.89961 \\
& 0.15 & 7.74806 & 7.69782 & 7.61304 & 7.49201 & 7.33209 \\
& 0.20 & 8.23336 & 8.18314 & 8.09846 & 7.97773 & 7.81848 \\
\hline
\multirow{4}{*}{$0.1$}
& 0.05 & 5.65931 & 5.58365 & 5.45417 & 5.26494 & 5.00525 \\
& 0.10 & 5.97510 & 5.89951 & 5.77033 & 5.58203 & 5.32478 \\
& 0.15 & 6.32805 & 6.25251 & 6.12364 & 5.93626 & 5.68137 \\
& 0.20 & 6.72511 & 6.64964 & 6.52106 & 6.33459 & 6.08199 \\
\hline
\end{tabular}
\end{minipage}

\label{tab:isco-radius}
\end{table*}

Both expressions diverge when $2f(r)-rf'(r)\to 0$, which is the condition that
defines the photon sphere; this confirms that the massive-particle circular orbits
merge smoothly into null circular orbits in the ultra-relativistic limit. The
denominator $2f-rf'$ must be positive for the orbit to be physically accessible.

The radial profiles of $\mathcal{L}_{\rm sp}$ and $\mathcal{E}_{\rm sp}$ are
presented in Fig.~\ref{fig:LspEsp}. Both quantities diverge as $r$ approaches the
photon sphere from above, and reach a local minimum at the ISCO, below which no
stable circular orbit exists. The upper row displays $\mathcal{L}_{\rm sp}$:
increasing $\alpha$ [panel~(a)] and decreasing $\ell$ [panel~(b)] shift the ISCO
outward and increase the angular momentum required for circular motion, while larger
charges $Q$ [panel~(c)] reduce $\mathcal{L}_{\rm sp}$ at a given radius. The ModMax
parameter $\gamma$ [panel~(d)] moves the curves toward the Schwarzschild profile as
$\gamma$ increases. The lower row shows $\mathcal{E}_{\rm sp}$: the local minimum
provides the binding energy of the ISCO, and the radiative efficiency of an accretion disc is $\eta=1-\mathcal{E}_{\rm sp}^{\rm ISCO}$; its dependence on the parameters can be read directly from panels~(e)--(h).

A circular orbit is stable against small radial perturbations if and only if it
corresponds to a local minimum of $U_{\rm eff}$. The three stability conditions
are therefore
\begin{align}
    &\mathcal{E}^2=\left(1+\frac{\mathcal{L}^2}{r^2}\right)f(r),\nonumber\\
    &\frac{\partial U_{\rm eff}}{\partial r}=0,\nonumber\\
    &\frac{\partial^2 U_{\rm eff}}{\partial r^2}\geq 0.\label{dd7}
\end{align}

The first two conditions locate the circular orbit, while the third ensures its stability: $\partial^2 U_{\rm eff}/\partial r^2>0$ corresponds to a restoring
force after radial perturbations, whereas $\partial^2 U_{\rm eff}/\partial r^2<0$ signals instability. The ISCO is defined by the marginally stable case
$\partial^2 U_{\rm eff}/\partial r^2=0$, which, when combined with Eqs.~\eqref{dd4}--\eqref{dd5}, yields the condition
\begin{equation}
    f(r)\,f''(r)-2\bigl[f'(r)\bigr]^2+\frac{3\,f(r)\,f'(r)}{r}=0.\label{dd8}
\end{equation}
The ISCO radius is found by solving Eq.~\eqref{dd8} numerically for $r$ given the metric function~\eqref{aa2}. For the Schwarzschild black hole ($\alpha=\ell=\gamma =Q_e=Q_m=0$), this equation reduces to $r=6M$, as expected. In Table \ref{tab:isco-radius}, we present numerical values of the ISCO radius by varying both the electric and magnetic charges, KR-field parameter $\ell$ and string cloud parameter $\alpha$ for two values of ModMax parameter $\gamma$.

\section{Thermodynamics}\label{sec:thermo}

In this section, we study the thermodynamics of the black hole solution~\eqref{aa2} and analyze how the model parameters alter the Hawking temperature, the entropy, and the specific heat capacity of the system.

Black hole thermodynamics plays a central role in the search for a consistent theory of quantum gravity, as it reveals a profound connection between general relativity, quantum mechanics, and statistical physics. The thermodynamic interpretation of black holes emerged in the early 1970s through the pioneering work of Bekenstein and Hawking. In particular, Bekenstein proposed that black holes should possess an entropy proportional to the area of their event horizon \cite{Bekestein1972, Bekestein1973}. This idea was further developed within the framework of the laws of black hole mechanics formulated by Bardeen et al. \cite{Bardeen1973}. The subject gained a solid physical foundation when Hawking demonstrated that black holes emit thermal radiation due to quantum effects near the event horizon, now known as Hawking radiation \cite{Hawking1975}. This discovery established that black holes can be consistently described as thermodynamic systems characterized by temperature, entropy, and energy. 

We begin by determining the horizon structure. Setting $f(r_h)=0$ in
Eq.~\eqref{aa2} and solving for $r_h$, we find the two horizons (outer $r_+$ and
inner $r_-$)
\begin{equation}
    r_{\pm}=\frac{1}{\lambda}\left(M\pm\sqrt{M^2-\lambda\,\zeta}\right),\label{ff0}
\end{equation}
where the shorthand parameters are defined as
\begin{equation}
    \lambda\equiv\frac{1-\alpha}{1-\ell},\qquad
    \zeta\equiv e^{-\gamma}\,\frac{Q_e^2+Q_m^2}{(1-\ell)^2}.\label{ff1}
\end{equation}
The parameter $\lambda$ is the asymptotic value of the lapse function, encoding the
joint modification of the geometry by the string cloud and the KR field, while
$\zeta$ is the effective (ModMax-screened) charge parameter. Real horizons exist
only when the discriminant is non-negative, imposing the constraint
\begin{equation}
    M^2>e^{-\gamma}\,\frac{(Q_e^2+Q_m^2)(1-\alpha)}{(1-\ell)^3}.\label{ff_exist}
\end{equation}
When the bound~\eqref{ff_exist} is saturated, $r_+=r_-$ and the black hole is
extremal; below it, the singularity is naked. One can verify that the photon-sphere
existence condition~\eqref{bb_exist} is more stringent than~\eqref{ff_exist} by
a factor of $8/9$, so a black hole always possesses a photon sphere as long as it
possesses a horizon.

The Hawking temperature is related to the surface gravity $\kappa$ at the outer horizon by $T=\kappa/(2\pi)$ \cite{Wald1993,IyerWald1994}. For the static metric~\eqref{aa1}, the surface gravity is determined by the gradient of the Killing vector $\xi^\mu=(\partial/\partial t)^\mu$ normal to the horizon,
\begin{equation}
    \kappa=\frac{1}{2}\sqrt{-(\nabla^\mu\xi^\nu)(\nabla_\mu\xi_\nu)},\label{ff3}
\end{equation}
which for the metric $g_{tt}g_{rr}=-f^2$ simplifies to the familiar expression
\begin{equation}
    \kappa=\frac{1}{2}\,\frac{|\partial_r g_{tt}|}{\sqrt{-g_{tt}g_{rr}}}
    =\frac{1}{2}\,f'(r_h).\label{ff4}
\end{equation}
Evaluating $f'(r)$ at $r=r_h$ using Eq.~\eqref{aa2} and substituting into
$T=\kappa/(2\pi)$, the Hawking temperature reads
\begin{equation}
    T=\frac{1}{2\pi r_h}\left[\frac{M}{r_h}
    -e^{-\gamma}\frac{(Q_e^2+Q_m^2)}{(1-\ell)^2 r_h^2}\right].\label{ff6}
\end{equation}
The first term inside the bracket is the gravitational (Schwarzschild-like)
contribution, while the second is the electromagnetic correction: a larger dyonic
charge lowers the temperature, and in the extremal limit $r_+=r_-$, the temperature
vanishes identically, in accordance with the third law of black hole mechanics.

\begin{figure*}[ht]
    \centering
    \includegraphics[width=\textwidth]{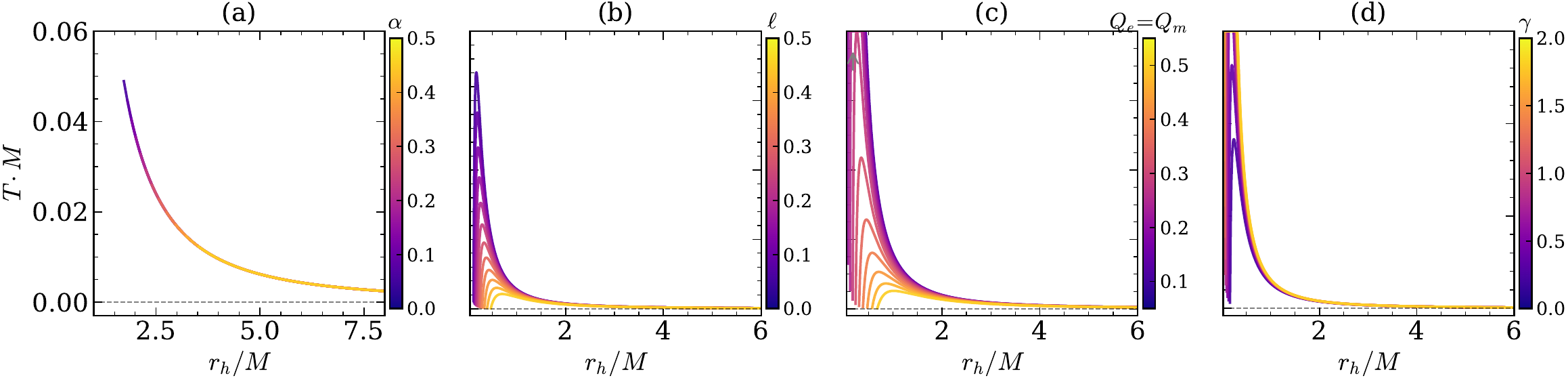}
    \caption{Hawking temperature $T\cdot M$ as a function of the horizon radius
    $r_h/M$, for $M=1$. Each panel shows 12 curves varying one parameter
    continuously (colormap from dark to light), while the remaining parameters are
    fixed at $\alpha=0.15$, $\ell=0.10$, $Q_e=0.30$, $Q_m=0.20$, $\gamma=0.10$.
    \textbf{(a)}~$\alpha\in[0.00,\,0.50]$: curves are plotted only for $r_h\geq
    r_+(\alpha)$, since $T$ does not depend on $\alpha$ explicitly.
    \textbf{(b)}~$\ell\in[0.00,\,0.50]$.
    \textbf{(c)}~$Q_e=Q_m\in[0.05,\,0.55]$.
    \textbf{(d)}~$\gamma\in[0.00,\,2.00]$.
    The dashed horizontal line marks $T=0$.}
    \label{fig:temp}
\end{figure*}

The behaviour of the Hawking temperature $T$ as a function of the horizon radius
$r_h$ is displayed in Fig.~\ref{fig:temp}. Before describing each panel
individually, it is important to note that Eq.~\eqref{ff6} contains no explicit
dependence on the string-cloud parameter $\alpha$: $T$ is a function of $r_h$,
$M$, $\gamma$, $\ell$, $Q_e$, and $Q_m$ only. For panels~(b)--(d), which span the
full range of $r_h$ starting just above the near-extremal radius $r_h\approx\zeta/M$,
the temperature displays the characteristic non-monotonic profile of charged black
holes: $T$ rises from zero at $r_h=\zeta/M$, reaches a maximum at
$r_h^*=3\zeta/(2M)$, and then decays as $\sim M/(2\pi r_h^2)$ for large $r_h$.
This behaviour reflects the competition between the gravitational term $M/r_h$ and
the electromagnetic term $\propto \zeta/r_h^2$ in Eq.~\eqref{ff6}: for $r_h<r_h^*$
the charge term dominates and suppresses $T$, while for $r_h>r_h^*$ the mass term
governs the decay.

Panel~(a) explores the role of $\alpha$ in the physically accessible domain. Since
$T$ is $\alpha$-independent, all curves share the same functional form $T(r_h)$;
what changes with $\alpha$ is the outer horizon radius $r_+(\alpha)$, which increases
as $\alpha$ grows. Each curve is therefore plotted only for $r_h\geq r_+(\alpha)$:
larger $\alpha$ shifts the starting point of the curve to a region where $T$ is
already on its descending branch and takes a lower value. This demonstrates that
the string cloud lowers the Hawking temperature of the physically realized black
hole by pushing the horizon outward, without altering the intrinsic $T(r_h)$ profile.
Panel~(b) shows the effect of $\ell$: increasing $\ell$ raises $\zeta\propto
(1-\ell)^{-2}$, which shifts the peak position $r_h^*=3\zeta/(2M)$ to larger radii
and simultaneously lowers the peak temperature $T_{\rm peak}\propto \zeta^{-2}$,
producing the family of lower and broader curves visible in the figure. Panel~(c)
shows that increasing the combined charge $Q_e=Q_m=Q$ has a qualitatively similar
effect: larger $Q$ raises $\zeta\propto Q^2$, pushing $r_h^*$ outward and sharply
suppressing $T_{\rm peak}$, so the spectrum of curves shifts toward lower and
broader profiles. The ModMax parameter $\gamma$ [panel~(d)] acts in the opposite
direction: larger $\gamma$ exponentially suppresses $\zeta\propto e^{-\gamma}$,
driving $r_h^*\to 0$ and $T_{\rm peak}\to\infty$, so the temperature peak migrates
to smaller radii and larger values; in the limit $\gamma\to\infty$ one recovers the
Schwarzschild result $T=1/(4\pi r_h)$, which decreases monotonically with no
finite maximum.

Assuming the system obeys the Bekenstein-Hawking area law, the entropy is
proportional to the area of the event horizon,
\begin{equation}
    S=\frac{\mathcal{A}}{4}=\pi r_h^2,\label{ff7}
\end{equation}
which is unaffected by the model parameters, since the horizon area is
$\mathcal{A}=4\pi r_h^2$ in the standard spherical topology. Combining
Eqs.~\eqref{ff6} and~\eqref{ff7}, one finds
\begin{equation}
    2TS=M-e^{-\gamma}\frac{(Q_e^2+Q_m^2)}{(1-\ell)^2 r_h}.\label{ff8}
\end{equation}

The black hole mass expressed as a function of the horizon radius follows from
$f(r_h)=0$,
\begin{equation}
    M=\frac{r_h}{2}\left[\frac{1-\alpha}{1-\ell}
    +e^{-\gamma}\frac{(Q_e^2+Q_m^2)}{(1-\ell)^2 r_h^2}\right].\label{ff9}
\end{equation}
Equation~\eqref{ff9} identifies the mass as the internal energy $M=M(S,Q_e,Q_m)$,
with $S$ determined by $r_h$ through Eq.~\eqref{ff7}. The first law of black hole
thermodynamics then follows from the total differential
\begin{align}
    dM&=\left(\frac{\partial M}{\partial S}\right)_{Q_e,Q_m}dS
      +\left(\frac{\partial M}{\partial Q_e}\right)_{S,Q_m}dQ_e
      +\left(\frac{\partial M}{\partial Q_m}\right)_{S,Q_e}dQ_m\nonumber\\
    &=T\,dS+\Phi_e\,dQ_e+\Phi_m\,dQ_m,\label{ff10}
\end{align}
where the electrostatic and magnetostatic potentials at the horizon are
\begin{align}
    \Phi_e &= e^{-\gamma}\frac{Q_e}{(1-\ell)^2 r_h},\label{ff11}\\
    \Phi_m &= e^{-\gamma}\frac{Q_m}{(1-\ell)^2 r_h}.\label{ff12}
\end{align}
Both potentials carry the ModMax screening factor $e^{-\gamma}$ and the KR
rescaling $(1-\ell)^{-2}$, consistent with the screened effective charges already
seen in the metric. The thermodynamic variables satisfy the generalized Smarr
mass formula
\begin{equation}
    2TS+\Phi_e\,Q_e+\Phi_m\,Q_m=M,\label{ff13}
\end{equation}
which can be verified by substituting Eqs.~\eqref{ff6}--\eqref{ff12} and using
Eq.~\eqref{ff9}. The Smarr relation is a consequence of the scaling properties of
the solution and generalizes the standard RN relation to include the effects of
the string cloud, the KR field, and the ModMax nonlinearity.

Finally, the stability of the black hole against Hawking evaporation is governed by
the specific heat at constant charges,
\begin{equation}
    C=T\,\left(\frac{dS}{dT}\right)
    =2\pi r_h^2\,\frac{M r_h-e^{-\gamma}\dfrac{(Q_e^2+Q_m^2)}{(1-\ell)^2}}
    {-2 M r_h+3\,e^{-\gamma}\dfrac{(Q_e^2+Q_m^2)}{(1-\ell)^2}}.\label{ff14}
\end{equation}
A positive specific heat ($C>0$) indicates a thermally stable black hole, while a
negative value signals instability. Inspecting the sign of Eq.~\eqref{ff14}: the
numerator $Mr_h-\zeta>0$ for all physically accessible radii $r_h>\zeta/M$, while
the denominator $-2Mr_h+3\zeta$ changes sign at $r_h^*=3\zeta/(2M)$.
Consequently, $C>0$ (thermally stable) for $r_h<r_h^*$ and $C<0$ (thermally
unstable) for $r_h>r_h^*$; the divergence at $r_h^*$ marks a Hawking-Page-type
phase transition between the thermally stable small branch and the unstable large
branch. For the reference parameter set, $r_h^*\approx 0.22\,M$ while the outer
horizon lies at $r_+\approx 2.04\,M$, so the physically realised black hole resides
entirely on the unstable ($C<0$) branch for $M=1$. For $\gamma\to\infty$ the
electromagnetic terms in Eqs.~\eqref{ff6} and~\eqref{ff14} vanish, recovering the
Schwarzschild result $T=1/(4\pi r_h)$ and $C=-2\pi r_h^2<0$ (always unstable).

\begin{figure*}[ht]
    \centering
    \includegraphics[width=\textwidth]{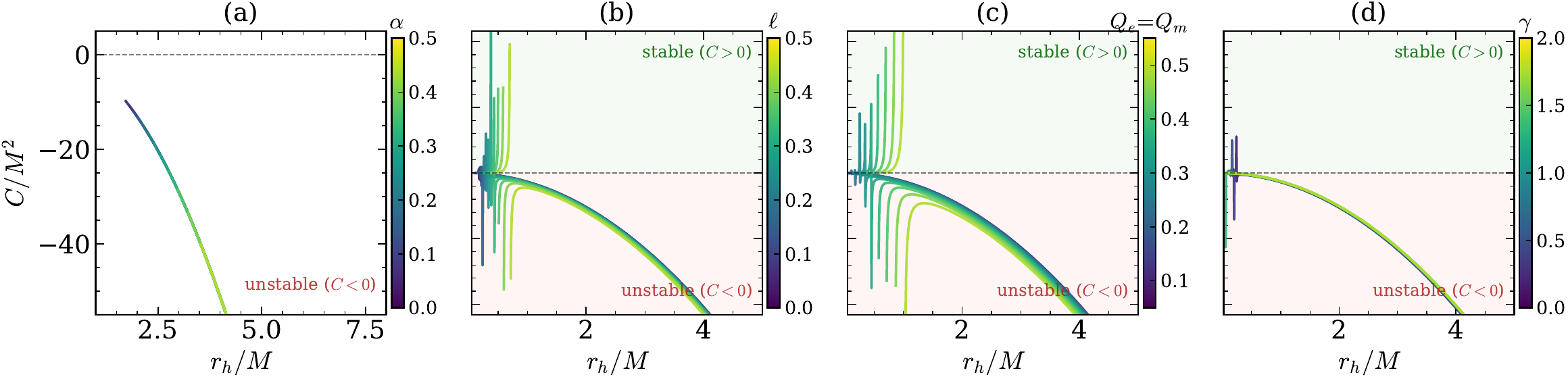}
    \caption{Specific heat $C/M^2$ as a function of the horizon radius $r_h/M$,
    for $M=1$. Each panel shows 12 curves varying one parameter continuously
    (colormap from dark to light), while the remaining parameters are fixed at
    $\alpha=0.15$, $\ell=0.10$, $Q_e=0.30$, $Q_m=0.20$, $\gamma=0.10$.
    \textbf{(a)}~$\alpha\in[0.00,\,0.50]$: curves are plotted only for $r_h\geq
    r_+(\alpha)$, since $C$ does not depend on $\alpha$ explicitly; all curves lie
    on the unstable ($C<0$) branch.
    \textbf{(b)}~$\ell\in[0.00,\,0.50]$.
    \textbf{(c)}~$Q_e=Q_m\in[0.05,\,0.55]$.
    \textbf{(d)}~$\gamma\in[0.00,\,2.00]$.
    The dashed horizontal line marks $C=0$; curves near the divergence at $r_h^*$
    have been clipped for clarity. Green (red) shading indicates the stable
    $C>0$ (unstable $C<0$) region.}
    \label{fig:sheat}
\end{figure*}

Figure~\ref{fig:sheat} shows $C$ as a function of $r_h$. As with the temperature,
Eq.~\eqref{ff14} contains no explicit dependence on $\alpha$, so the $C(r_h)$
profile is the same for all values of $\alpha$. Panel~(a) confirms this: since the
plotted range is restricted to $r_h\geq r_+(\alpha)\gg r_h^*$, every curve lies
entirely on the unstable ($C<0$) branch. As $\alpha$ increases, $r_+$ shifts
outward and the displayed portion of the curve extends to larger $r_h$, with $|C|$
growing approximately as $2\pi r_h^2$ far from the transition; this is the only
visible effect of $\alpha$ in this panel. Panels~(b)--(d) display the complete
$C(r_h)$ profile, starting just above the near-extremal radius, so that both
branches and the divergence at $r_h^*$ are visible. Panel~(b) shows that
increasing $\ell$ raises $\zeta\propto(1-\ell)^{-2}$, which pushes $r_h^*=
3\zeta/(2M)$ to larger values; this widens the stable ($C>0$) near-extremal branch
and shifts the Hawking-Page transition outward. Panel~(c) shows the analogous
effect of larger charges $Q$: $r_h^*\propto Q^2$ increases, broadening the stable
branch and delaying the onset of instability. Conversely, increasing $\gamma$
[panel~(d)] suppresses $\zeta\propto e^{-\gamma}$ and drives $r_h^*\to 0$,
shrinking the stable branch; in the limit $\gamma\to\infty$ one recovers
$C=-2\pi r_h^2<0$ everywhere, and the stable branch disappears entirely.

\section{Sparsity of Hawking radiation}\label{sec:sparsity}

In this section, we quantify the sparsity of Hawking radiation for our black hole solution. Although a black hole radiates thermally at a temperature determined by its surface gravity, the Hawking emission is temporally discrete, consisting of well-separated quanta rather than a continuous flux. Sparsity is conventionally characterized by comparing the square of the thermal wavelength \(\lambda_t=2\pi/T\) with the effective emission area \(\mathcal{A}_{\rm eff}\). Following the definition introduced in \cite{Hawking1975, Page1976, Gray2016}, the dimensionless sparsity parameter is given by
\begin{equation}\label{Sparsity}
    \eta =\frac{\mathcal{C}}{\Tilde{g} }\left(\frac{\lambda_t^2}{\mathcal{A}_{\rm eff}}\right),
\end{equation}
where \(\mathcal{C}\) is a dimensionless constant, \(\tilde g\) the spin degeneracy of the emitted quanta and  \(\mathcal{A}_{\rm eff}=\frac{27}{4}\,A_{\rm BH}=27\pi r_h^2\), where the horizon area in our case is $A_{\rm BH}=4 \pi r_h^2$. 

For our KR- gravity sourced by a cloud of string for ModMax dyonic black hole solution, the modified surface gravity and horizon geometry influence both the Hawking temperature and the effective emission area. Consequently, the sparsity parameter exhibits a nontrivial dependence on the parameters $(Q_e,\,Q_m,\,\ell,\,\gamma,\,\alpha)$, directly affecting the discreteness and observational signatures of the evaporation process. 
Substituting the temperature \eqref{ff6} into the sparsity definition \eqref{Sparsity} yields the following closed form
\begin{equation}\label{sspp}
\eta=\frac{16 \pi ^3}{27 \left(\frac{M}{r_h}-\frac{\zeta}{r_h^2} \right)^2},
\end{equation}
where the horizon $r_h=r_{+}$ is in (\ref{ff0}) and its expression is given by
\begin{equation}
    r_h=\frac{1}{\lambda}\left(M+\sqrt{M^2-\lambda\,\zeta}\right).\label{horizon-radius}
\end{equation}

\section{Rate of Energy Emission}\label{sec:emission}

Quantum effects in curved spacetime imply that black holes are not perfectly black:
a thermal flux is emitted from the near-horizon region with a temperature fixed by
the surface gravity. In the semiclassical language of particle-pair production near
the event horizon, this Hawking radiation carries energy to infinity and leads to a
gradual decrease of the black hole mass~\cite{Javed2019}. From the perspective of a
distant observer, the radiated power depends not only on the Hawking temperature but
also on the probability that emitted quanta successfully cross the gravitational
potential barrier between the horizon and infinity.

In the geometric-optics (high-frequency) limit, the absorption cross-section
oscillates around the constant value~\cite{Misner1973,Mashhoon1973,Wei2013}
\begin{equation}
    \sigma_{\rm lim}\approx\pi R_{\rm sh}^2,\label{ee1}
\end{equation}
where $R_{\rm sh}$ is the black hole shadow radius given in Eq.~\eqref{bb12}.
The identification $\sigma_{\rm lim}\propto R_{\rm sh}^2$ has a clear physical
origin: at high energies, the dominant contribution to the absorption cross section
comes from photons (and massless particles) whose impact parameter equals the
critical value $\beta_c$, i.e.\ those that spiral asymptotically onto the photon
sphere. Thus, the shadow radius acts as the effective size of the absorber for
high-frequency radiation, directly linking the optical and thermodynamic properties
of the black hole.

Within this geometric-optics approximation, the spectral energy emission rate is
given by a grey-body-corrected Planck spectrum~\cite{Wei2013},
\begin{equation}
    \frac{d^2\mathbb{E}}{d\omega\,dt}
    =\frac{2\pi^2\,\sigma_{\rm lim}}{e^{\omega/T}-1}\,\omega^3,\label{ee2}
\end{equation}
where $\omega$ is the emitted frequency and $T$ is the Hawking temperature given in
Eq.~\eqref{ff6}. The factor $(e^{\omega/T}-1)^{-1}$ is the Bose-Einstein occupation
number of the thermal radiation, while $\omega^3$ comes from the density of states
in three spatial dimensions. Substituting Eqs.~\eqref{bb12} and~\eqref{ff6}
into Eq.~\eqref{ee2}, the full emission rate for a finite observer at $r_O$ is
\begin{align}
    \frac{d^2\mathbb{E}}{d\omega\,dt}
    &=2\pi^3\omega^3\,r_s^2
    \left(\frac{f(r_O)}{f(r_s)}\right)
    \nonumber\\[4pt]
    &\quad\times\left[\exp\!\left\{
    \frac{2\pi r_h\,\omega}
    {\dfrac{M}{r_h}-e^{-\gamma}\dfrac{Q_e^2+Q_m^2}{(1-\ell)^2 r_h^2}}
    \right\}-1\right]^{-1},\label{ee3}
\end{align}
where $r_s$ is given by Eq.~\eqref{bb8}, and $f(r_O)$, $f(r_s)$ denote the lapse
function~\eqref{aa2} evaluated at $r_O$ and $r_s$ respectively. The ratio
$f(r_O)/f(r_s)$ is the finite-distance correction to the shadow area that vanishes
in the flat-background limit. In the distant-observer limit $r_O\to\infty$,
$f(r_O)\to(1-\alpha)/(1-\ell)$, and the expression simplifies to
\begin{align}
    \frac{d^2\mathbb{E}}{d\omega\,dt}
    &=\frac{1-\alpha}{1-\ell}\cdot
    \frac{2\pi^3\omega^3\,r_s^2}{f(r_s)}
    \nonumber\\[4pt]
    &\quad\times\left[\exp\!\left\{
    \frac{2\pi r_h\,\omega}
    {\dfrac{M}{r_h}-e^{-\gamma}\dfrac{Q_e^2+Q_m^2}{(1-\ell)^2 r_h^2}}
    \right\}-1\right]^{-1}.\label{ee4}
\end{align}
The prefactor $(1-\alpha)/(1-\ell)$ in Eq.~\eqref{ee4} is precisely the asymptotic
value of the lapse function $f_\infty$ introduced earlier; it amplifies or suppresses
the overall emission power depending on whether the KR field dominates ($\ell>\alpha$)
or the string cloud dominates ($\alpha>\ell$).

\begin{figure*}[ht]
    \centering
    \includegraphics[width=\textwidth]{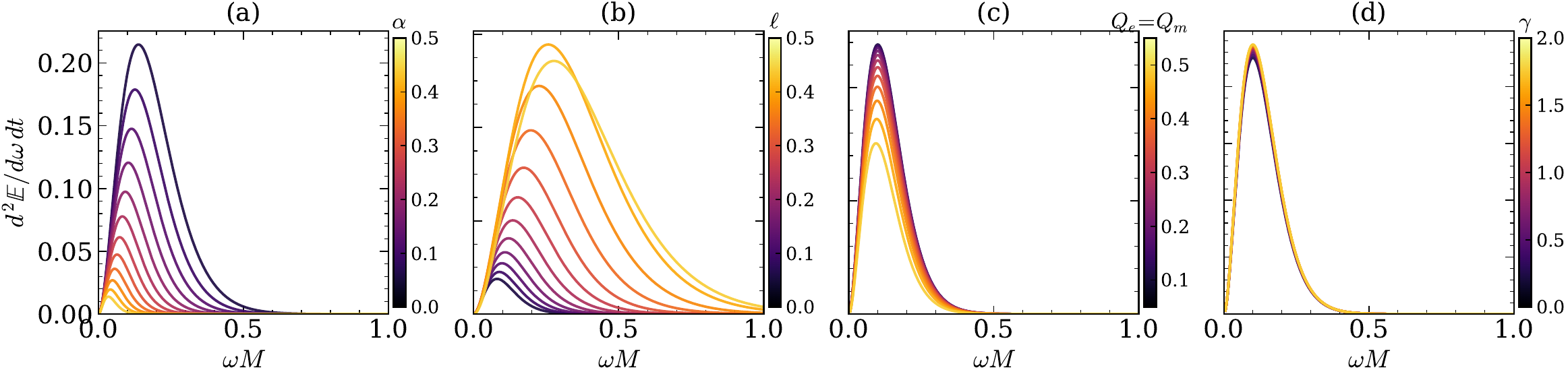}
    \caption{Spectral energy emission rate $d^2\mathbb{E}/d\omega\,dt$ as a function
    of the emission frequency $\omega M$, for a distant observer and $M=1$. Each
    panel shows 12 curves varying one parameter continuously (colormap from dark to
    light), while the remaining parameters are fixed at $\alpha=0.15$, $\ell=0.10$,
    $Q_e=0.30$, $Q_m=0.20$, $\gamma=0.10$.
    \textbf{(a)}~$\alpha\in[0.00,\,0.50]$.
    \textbf{(b)}~$\ell\in[0.00,\,0.50]$.
    \textbf{(c)}~$Q_e=Q_m\in[0.05,\,0.55]$.
    \textbf{(d)}~$\gamma\in[0.00,\,2.00]$.}
    \label{fig:emission}
\end{figure*}

Figure~\ref{fig:emission} displays the spectral energy emission rate as a function
of the emission frequency $\omega$ for a distant observer. The emission spectrum has
the characteristic shape of a modified Planck distribution: it vanishes at $\omega=0$,
rises steeply, reaches a peak at a frequency $\omega_{\rm peak}\approx 2.82\,T$
set by Wien's displacement law, and then falls off exponentially for $\omega\gg T$.
Any modification that raises the Hawking temperature therefore shifts the peak to
higher frequencies and increases the total emitted power. Panel~(a) shows that
increasing $\alpha$ reduces the prefactor $(1-\alpha)/(1-\ell)$ in Eq.~\eqref{ee4}
and lowers the overall amplitude, producing a dimmer spectrum; since $\alpha$ does
not enter the Hawking temperature directly, the peak frequency is not significantly
shifted. Panel~(b) shows the complementary effect of $\ell$: larger $\ell$ increases
the prefactor $(1-\alpha)/(1-\ell)$ and enhances the emission amplitude, while
simultaneously lowering the Hawking temperature at the physical horizon
(cf.\ Fig.~\ref{fig:temp}), resulting in a brighter but slightly cooler spectrum.
Larger combined charges $Q$ [panel~(c)] substantially lower the Hawking temperature
(cf.\ Fig.~\ref{fig:temp}), so the peak migrates to lower frequencies and the total
emission power drops sharply. Finally, increasing $\gamma$ [panel~(d)] exponentially
suppresses the effective charge $e^{-\gamma}Q^2$, which raises the temperature at
the physical horizon and shifts the emission peak to higher frequencies and larger
amplitudes; in the limit $\gamma\to\infty$ the spectrum approaches the Schwarzschild
result.

The energy emission rate is thus influenced by all five parameters of the model:
the ModMax parameter $\gamma$ reduces the effective electromagnetic source
(decreasing both the photon sphere size and the Hawking temperature correction),
the dyonic charges $Q_e$ and $Q_m$ affect the photon sphere radius $r_s$ and the
horizon temperature in competing ways, while the string cloud parameter $\alpha$
and the KR-field parameter $\ell$ set the overall scale of the emission through
the conical-geometry prefactor.

\section{Summary and Conclusions}\label{sec:conclusions}

In this work, we have investigated the geodesic structure, thermodynamics, and
energy emission of a dyonic ModMax black hole embedded in a Kalb-Ramond gravity
background with a cloud of strings. The lapse function of the solution contains
five physical parameters beyond the mass $M$: the electric and magnetic charges
$Q_e$ and $Q_m$, the ModMax nonlinearity parameter $\gamma$, the KR-field parameter
$\ell$, and the string-cloud parameter $\alpha$. The solution interpolates smoothly
between well-known limits, Reissner-Nordstr\"{o}m for $\alpha=\ell=\gamma=0$,
Schwarzschild, when the charges also vanish, and the KR black hole of Ref.~\cite{Yang2023} when $\gamma=\alpha=0$.

A key structural feature of the solution is the non-flat asymptotics: both the
cloud of strings and the KR field introduce a conical singularity at spatial
infinity, so the metric approaches the constant $f_\infty=(1-\alpha)/(1-\ell)\neq 1$ rather than unity. This global modification propagates into all observables. The shadow radius of a distant observer acquires the multiplicative factor
$\sqrt{(1-\alpha)/(1-\ell)}$ relative to the critical impact parameter, and the
spectral energy emission rate is rescaled by the same quantity, neither of which can be reproduced by a simple redefinition of the mass or charges. The ModMax parameter $\gamma$ plays a complementary role: it exponentially screens the effective dyonic charge through the factor $e^{-\gamma}$, providing a one-parameter family of solutions that interpolate continuously between the full dyonic RN case ($\gamma=0$) and the uncharged Schwarzschild limit ($\gamma\to\infty$), with all photon-sphere, shadow, and thermodynamic quantities varying monotonically along this interpolation.

The photon-sphere analysis showed that both the photon sphere radius $r_s$ and the
critical impact parameter $\beta_c$ decrease with increasing dyonic charge, as in the standard RN spacetime, and increase as $\gamma$ grows, and the electromagnetic contribution is suppressed. The photon sphere exists only when the
mass-to-charge ratio satisfies the bound~\eqref{bb_exist}, which is more stringent
than the horizon existence bound~\eqref{ff_exist} by a factor of $8/9$; consequently, a horizon always implies a photon sphere in this spacetime. The effective radial force analysis confirmed that the photon sphere is an unstable equilibrium, photons inside $r_s$ experience a net inward pull while those outside experience an outward push, and that the zero of the force is shifted outward by $\alpha$ and inward by increasing charge, in full consistency with the photon sphere results.

For massive test particles, the rest-mass term in the effective potential generates
a local minimum absent in the photon case, signalling the existence of stable circular orbits. The ISCO, located by the vanishing of $\partial^2 U_{\rm eff}/\partial r^2$, migrates outward and requires higher specific angular momentum as $\alpha$ increases, while larger $\gamma$ drives the ISCO toward the Schwarzschild value $r_{\rm ISCO}=6M$. The accretion efficiency $\eta=1-\mathcal{E}_{\rm sp}^{\rm ISCO}$ is therefore sensitive to all five parameters, with the string cloud and KR field modifying it through the global geometry rather than through the electromagnetic sector.

On the thermodynamic side, the Hawking temperature decreases monotonically with
increasing dyonic charge and vanishes at the extremal limit, in accordance with the
third law. The entropy obeys the standard Bekenstein-Hawking area law $S=\pi r_h^2$
and is independent of the model parameters at a fixed horizon radius. The electric and magnetic potentials both carry the ModMax screening factor $e^{-\gamma}$ and the KR rescaling $(1-\ell)^{-2}$, and together with the temperature, they satisfy the first law and the generalized Smarr relation. The specific heat changes sign at a critical horizon radius, indicating a Hawking-Page-type phase transition: large black holes are thermally stable ($C>0$) while small ones are unstable ($C<0$), a behavior that persists for all values of the model parameters.

The Hawking emission rate connects the optical and thermodynamic sectors: the geometric-optics absorption cross section $\sigma_{\rm lim}\approx\pi R_{\rm sh}^2$ links the shadow radius directly to the spectral emissivity, so every parameter that shrinks the shadow also reduces the emission power, and vice versa. For a distant observer, the emission rate acquires the same conical prefactor
$(1-\alpha)/(1-\ell)$ that governs the shadow, so the ratio of emission rates in
two configurations with different $\alpha$ or $\ell$ values is determined entirely
by the global geometry, independently of the detailed horizon structure.

Taken together, these results demonstrate that the combined presence of ModMax nonlinear electrodynamics, the Kalb-Ramond background, and the cloud of strings produces a rich and internally consistent phenomenology. Future work will address the quasinormal mode spectrum of this geometry, the weak gravitational lensing angle via the Gauss-Bonnet theorem, and a quantitative comparison of the predicted
shadow size with the EHT constraints on M87* and Sgr A*.

\section*{Acknowledgments}

F.A. acknowledges the Inter University Centre for Astronomy and Astrophysics (IUCAA), Pune, India, for granting a visiting associateship. E. O. Silva acknowledges the support from grants CNPq/306308/2022-3, FAPEMA/UNIVERSAL-06395/22, FAPEMA/APP-12256/22, and (CAPES) - Brazil (Code 001).

\section*{Data Availability Statement}

No new data were generated or analyzed in this manuscript.

\section*{Conflict of Interest}

Author(s) declares no conflict of interest in this manuscript.


\begin{thebibliography}{49}%
	\makeatletter
	\providecommand \@ifxundefined [1]{%
		\@ifx{#1\undefined}
	}%
	\providecommand \@ifnum [1]{%
		\ifnum #1\expandafter \@firstoftwo
		\else \expandafter \@secondoftwo
		\fi
	}%
	\providecommand \@ifx [1]{%
		\ifx #1\expandafter \@firstoftwo
		\else \expandafter \@secondoftwo
		\fi
	}%
	\providecommand \natexlab [1]{#1}%
	\providecommand \enquote  [1]{``#1''}%
	\providecommand \bibnamefont  [1]{#1}%
	\providecommand \bibfnamefont [1]{#1}%
	\providecommand \citenamefont [1]{#1}%
	\providecommand \href@noop [0]{\@secondoftwo}%
	\providecommand \href [0]{\begingroup \@sanitize@url \@href}%
	\providecommand \@href[1]{\@@startlink{#1}\@@href}%
	\providecommand \@@href[1]{\endgroup#1\@@endlink}%
	\providecommand \@sanitize@url [0]{\catcode `\\12\catcode `\$12\catcode
		`\&12\catcode `\#12\catcode `\^12\catcode `\_12\catcode `\%12\relax}%
	\providecommand \@@startlink[1]{}%
	\providecommand \@@endlink[0]{}%
	\providecommand \url  [0]{\begingroup\@sanitize@url \@url }%
	\providecommand \@url [1]{\endgroup\@href {#1}{\urlprefix }}%
	\providecommand \urlprefix  [0]{URL }%
	\providecommand \Eprint [0]{\href }%
	\providecommand \doibase [0]{https://doi.org/}%
	\providecommand \selectlanguage [0]{\@gobble}%
	\providecommand \bibinfo  [0]{\@secondoftwo}%
	\providecommand \bibfield  [0]{\@secondoftwo}%
	\providecommand \translation [1]{[#1]}%
	\providecommand \BibitemOpen [0]{}%
	\providecommand \bibitemStop [0]{}%
	\providecommand \bibitemNoStop [0]{.\EOS\space}%
	\providecommand \EOS [0]{\spacefactor3000\relax}%
	\providecommand \BibitemShut  [1]{\csname bibitem#1\endcsname}%
	\let\auto@bib@innerbib\@empty
	\bibitem [{\citenamefont {Akiyama}\ \emph
		{et~al.}(2019{\natexlab{a}})\citenamefont {Akiyama} \emph
		{et~al.}}]{EHT2019L1}%
	\BibitemOpen
	\bibfield  {author} {\bibinfo {author} {\bibfnamefont {K.}~\bibnamefont
			{Akiyama}} \emph {et~al.} (\bibinfo {collaboration} {Event Horizon
			Telescope}),\ }\href {https://doi.org/10.3847/2041-8213/ab0ec7} {\bibfield
		{journal} {\bibinfo  {journal} {Astrophys. J. Lett.}\ }\textbf {\bibinfo
			{volume} {875}},\ \bibinfo {pages} {L1} (\bibinfo {year}
		{2019}{\natexlab{a}})}\BibitemShut {NoStop}%
	\bibitem [{\citenamefont {Akiyama}\ and\ \citenamefont {others (Event Horizon
			Telescope~Collaboration)}(2019)}]{EHT2019L4}%
	\BibitemOpen
	\bibfield  {author} {\bibinfo {author} {\bibfnamefont {K.}~\bibnamefont
			{Akiyama}}\ and\ \bibinfo {author} {\bibnamefont {others (Event Horizon
				Telescope~Collaboration)}},\ }\href
	{https://doi.org/10.3847/2041-8213/ab0e85} {\bibfield  {journal} {\bibinfo
			{journal} {Astrophys.\ J.\ Lett.}\ }\textbf {\bibinfo {volume} {875}},\
		\bibinfo {pages} {L4} (\bibinfo {year} {2019})}\BibitemShut {NoStop}%
	\bibitem [{\citenamefont {Akiyama}\ \emph
		{et~al.}(2019{\natexlab{b}})\citenamefont {Akiyama} \emph
		{et~al.}}]{EHT2019L6}%
	\BibitemOpen
	\bibfield  {author} {\bibinfo {author} {\bibfnamefont {K.}~\bibnamefont
			{Akiyama}} \emph {et~al.} (\bibinfo {collaboration} {Event Horizon
			Telescope}),\ }\href {https://doi.org/10.3847/2041-8213/ab1141} {\bibfield
		{journal} {\bibinfo  {journal} {Astrophys. J. Lett.}\ }\textbf {\bibinfo
			{volume} {875}},\ \bibinfo {pages} {L6} (\bibinfo {year}
		{2019}{\natexlab{b}})}\BibitemShut {NoStop}%
	\bibitem [{\citenamefont {Akiyama}\ \emph
		{et~al.}(2022{\natexlab{a}})\citenamefont {Akiyama} \emph
		{et~al.}}]{EHT2022L12}%
	\BibitemOpen
	\bibfield  {author} {\bibinfo {author} {\bibfnamefont {K.}~\bibnamefont
			{Akiyama}} \emph {et~al.} (\bibinfo {collaboration} {Event Horizon
			Telescope}),\ }\href {https://doi.org/10.3847/2041-8213/ac6674} {\bibfield
		{journal} {\bibinfo  {journal} {Astrophys. J. Lett.}\ }\textbf {\bibinfo
			{volume} {930}},\ \bibinfo {pages} {L12} (\bibinfo {year}
		{2022}{\natexlab{a}})}\BibitemShut {NoStop}%
	\bibitem [{\citenamefont {Akiyama}\ \emph
		{et~al.}(2022{\natexlab{b}})\citenamefont {Akiyama} \emph
		{et~al.}}]{EHT2022L17}%
	\BibitemOpen
	\bibfield  {author} {\bibinfo {author} {\bibfnamefont {K.}~\bibnamefont
			{Akiyama}} \emph {et~al.} (\bibinfo {collaboration} {Event Horizon
			Telescope}),\ }\href {https://doi.org/10.3847/2041-8213/ac6756} {\bibfield
		{journal} {\bibinfo  {journal} {Astrophys. J. Lett.}\ }\textbf {\bibinfo
			{volume} {930}},\ \bibinfo {pages} {L17} (\bibinfo {year}
		{2022}{\natexlab{b}})}\BibitemShut {NoStop}%
	\bibitem [{\citenamefont {Vagnozzi}\ \emph {et~al.}(2023)\citenamefont
		{Vagnozzi} \emph {et~al.}}]{Vagnozzi2023}%
	\BibitemOpen
	\bibfield  {author} {\bibinfo {author} {\bibfnamefont {S.}~\bibnamefont
			{Vagnozzi}} \emph {et~al.},\ }\href
	{https://doi.org/10.1088/1361-6382/acd97b} {\bibfield  {journal} {\bibinfo
			{journal} {Class. Quant. Grav.}\ }\textbf {\bibinfo {volume} {40}},\ \bibinfo
		{pages} {165007} (\bibinfo {year} {2023})}\BibitemShut {NoStop}%
	\bibitem [{\citenamefont {Born}\ and\ \citenamefont
		{Infeld}(1934)}]{BornInfeld1934}%
	\BibitemOpen
	\bibfield  {author} {\bibinfo {author} {\bibfnamefont {M.}~\bibnamefont
			{Born}}\ and\ \bibinfo {author} {\bibfnamefont {L.}~\bibnamefont {Infeld}},\
	}\href {https://doi.org/10.1098/rspa.1934.0059} {\bibfield  {journal}
		{\bibinfo  {journal} {Proc. Roy. Soc. Lond. A}\ }\textbf {\bibinfo {volume}
			{144}},\ \bibinfo {pages} {425} (\bibinfo {year} {1934})}\BibitemShut
	{NoStop}%
	\bibitem [{\citenamefont {Bandos}\ \emph {et~al.}(2020)\citenamefont {Bandos},
		\citenamefont {Lechner}, \citenamefont {Sorokin},\ and\ \citenamefont
		{Townsend}}]{Bandos2020}%
	\BibitemOpen
	\bibfield  {author} {\bibinfo {author} {\bibfnamefont {I.}~\bibnamefont
			{Bandos}}, \bibinfo {author} {\bibfnamefont {K.}~\bibnamefont {Lechner}},
		\bibinfo {author} {\bibfnamefont {D.}~\bibnamefont {Sorokin}},\ and\ \bibinfo
		{author} {\bibfnamefont {P.~K.}\ \bibnamefont {Townsend}},\ }\href
	{https://doi.org/10.1103/PhysRevD.102.121703} {\bibfield  {journal} {\bibinfo
			{journal} {Phys. Rev. D}\ }\textbf {\bibinfo {volume} {102}},\ \bibinfo
		{pages} {121703} (\bibinfo {year} {2020})}\BibitemShut {NoStop}%
	\bibitem [{\citenamefont {Kosyakov}(2020)}]{Kosyakov2020}%
	\BibitemOpen
	\bibfield  {author} {\bibinfo {author} {\bibfnamefont {B.~P.}\ \bibnamefont
			{Kosyakov}},\ }\href {https://doi.org/10.1016/j.physletb.2020.135840}
	{\bibfield  {journal} {\bibinfo  {journal} {Phys. Lett. B}\ }\textbf
		{\bibinfo {volume} {810}},\ \bibinfo {pages} {135840} (\bibinfo {year}
		{2020})}\BibitemShut {NoStop}%
	\bibitem [{\citenamefont {Bandos}\ \emph {et~al.}(2021)\citenamefont {Bandos},
		\citenamefont {Lechner}, \citenamefont {Sorokin},\ and\ \citenamefont
		{Townsend}}]{Bandos2021Susy}%
	\BibitemOpen
	\bibfield  {author} {\bibinfo {author} {\bibfnamefont {I.}~\bibnamefont
			{Bandos}}, \bibinfo {author} {\bibfnamefont {K.}~\bibnamefont {Lechner}},
		\bibinfo {author} {\bibfnamefont {D.}~\bibnamefont {Sorokin}},\ and\ \bibinfo
		{author} {\bibfnamefont {P.~K.}\ \bibnamefont {Townsend}},\ }\href
	{https://doi.org/10.1007/JHEP10(2021)031} {\bibfield  {journal} {\bibinfo
			{journal} {JHEP}\ }\textbf {\bibinfo {volume} {2021}}\bibinfo  {number} {
			(10)},\ \bibinfo {pages} {31}}\BibitemShut {NoStop}%
	\bibitem [{\citenamefont {Babaei-Aghbolagh}\ \emph {et~al.}(2022)\citenamefont
		{Babaei-Aghbolagh}, \citenamefont {Velni}, \citenamefont {Yekta},\ and\
		\citenamefont {Mohammadzadeh}}]{Babaei2022}%
	\BibitemOpen
	\bibfield  {number} {  }\bibfield  {author} {\bibinfo {author} {\bibfnamefont
			{H.}~\bibnamefont {Babaei-Aghbolagh}}, \bibinfo {author} {\bibfnamefont
			{K.~B.}\ \bibnamefont {Velni}}, \bibinfo {author} {\bibfnamefont {D.~M.}\
			\bibnamefont {Yekta}},\ and\ \bibinfo {author} {\bibfnamefont
			{H.}~\bibnamefont {Mohammadzadeh}},\ }\href
	{https://doi.org/10.1016/j.physletb.2022.137079} {\bibfield  {journal}
		{\bibinfo  {journal} {Phys. Lett. B}\ }\textbf {\bibinfo {volume} {829}},\
		\bibinfo {pages} {137079} (\bibinfo {year} {2022})}\BibitemShut {NoStop}%
	\bibitem [{\citenamefont {Lechner}\ \emph {et~al.}(2022)\citenamefont
		{Lechner}, \citenamefont {Marchetti}, \citenamefont {Sainaghi},\ and\
		\citenamefont {Sorokin}}]{Lechner2022}%
	\BibitemOpen
	\bibfield  {author} {\bibinfo {author} {\bibfnamefont {K.}~\bibnamefont
			{Lechner}}, \bibinfo {author} {\bibfnamefont {P.}~\bibnamefont {Marchetti}},
		\bibinfo {author} {\bibfnamefont {A.}~\bibnamefont {Sainaghi}},\ and\
		\bibinfo {author} {\bibfnamefont {D.~P.}\ \bibnamefont {Sorokin}},\ }\href
	{https://doi.org/10.1103/PhysRevD.106.016009} {\bibfield  {journal} {\bibinfo
			{journal} {Phys. Rev. D}\ }\textbf {\bibinfo {volume} {106}},\ \bibinfo
		{pages} {016009} (\bibinfo {year} {2022})}\BibitemShut {NoStop}%
	\bibitem [{\citenamefont {Kruglov}(2022)}]{Kruglov2022}%
	\BibitemOpen
	\bibfield  {author} {\bibinfo {author} {\bibfnamefont {S.~I.}\ \bibnamefont
			{Kruglov}},\ }\href {https://doi.org/10.1142/S0218271822500250} {\bibfield
		{journal} {\bibinfo  {journal} {Int. J. Mod. Phys. D}\ }\textbf {\bibinfo
			{volume} {31}},\ \bibinfo {pages} {2250025} (\bibinfo {year}
		{2022})}\BibitemShut {NoStop}%
	\bibitem [{\citenamefont {Flores-Alfonso}\ \emph {et~al.}(2021)\citenamefont
		{Flores-Alfonso}, \citenamefont {Gonz\'{a}lez-Morales}, \citenamefont
		{Linares},\ and\ \citenamefont {Maceda}}]{FloresAlfonso2021}%
	\BibitemOpen
	\bibfield  {author} {\bibinfo {author} {\bibfnamefont {D.}~\bibnamefont
			{Flores-Alfonso}}, \bibinfo {author} {\bibfnamefont {B.~A.}\ \bibnamefont
			{Gonz\'{a}lez-Morales}}, \bibinfo {author} {\bibfnamefont {R.}~\bibnamefont
			{Linares}},\ and\ \bibinfo {author} {\bibfnamefont {M.}~\bibnamefont
			{Maceda}},\ }\href {https://doi.org/10.1016/j.physletb.2020.136011}
	{\bibfield  {journal} {\bibinfo  {journal} {Phys. Lett. B}\ }\textbf
		{\bibinfo {volume} {812}},\ \bibinfo {pages} {136011} (\bibinfo {year}
		{2021})}\BibitemShut {NoStop}%
	\bibitem [{\citenamefont {Pantig}\ \emph {et~al.}(2022)\citenamefont {Pantig},
		\citenamefont {Mastrototaro}, \citenamefont {Lambiase},\ and\ \citenamefont
		{\"{O}vg\"{u}n}}]{Pantig2022}%
	\BibitemOpen
	\bibfield  {author} {\bibinfo {author} {\bibfnamefont {R.~C.}\ \bibnamefont
			{Pantig}}, \bibinfo {author} {\bibfnamefont {L.}~\bibnamefont
			{Mastrototaro}}, \bibinfo {author} {\bibfnamefont {G.}~\bibnamefont
			{Lambiase}},\ and\ \bibinfo {author} {\bibfnamefont {A.}~\bibnamefont
			{\"{O}vg\"{u}n}},\ }\href {https://doi.org/10.1140/epjc/s10052-022-11125-y}
	{\bibfield  {journal} {\bibinfo  {journal} {Eur. Phys. J. C}\ }\textbf
		{\bibinfo {volume} {82}},\ \bibinfo {pages} {1155} (\bibinfo {year}
		{2022})}\BibitemShut {NoStop}%
	\bibitem [{\citenamefont {Barrientos}\ \emph {et~al.}(2022)\citenamefont
		{Barrientos}, \citenamefont {Cisterna}, \citenamefont {Kubiz\v{n}\'{a}k},\
		and\ \citenamefont {Oliva}}]{Barrientos2022}%
	\BibitemOpen
	\bibfield  {author} {\bibinfo {author} {\bibfnamefont {J.}~\bibnamefont
			{Barrientos}}, \bibinfo {author} {\bibfnamefont {A.}~\bibnamefont
			{Cisterna}}, \bibinfo {author} {\bibfnamefont {D.}~\bibnamefont
			{Kubiz\v{n}\'{a}k}},\ and\ \bibinfo {author} {\bibfnamefont {J.}~\bibnamefont
			{Oliva}},\ }\href {https://doi.org/10.1016/j.physletb.2022.137447} {\bibfield
		{journal} {\bibinfo  {journal} {Phys. Lett. B}\ }\textbf {\bibinfo {volume}
			{834}},\ \bibinfo {pages} {137447} (\bibinfo {year} {2022})}\BibitemShut
	{NoStop}%
	\bibitem [{\citenamefont {Ballon~Bordo}\ \emph {et~al.}(2021)\citenamefont
		{Ballon~Bordo}, \citenamefont {Kubiz\v{n}\'{a}k},\ and\ \citenamefont
		{Perche}}]{BallonBordo2021}%
	\BibitemOpen
	\bibfield  {author} {\bibinfo {author} {\bibfnamefont {A.}~\bibnamefont
			{Ballon~Bordo}}, \bibinfo {author} {\bibfnamefont {D.}~\bibnamefont
			{Kubiz\v{n}\'{a}k}},\ and\ \bibinfo {author} {\bibfnamefont {T.~R.}\
			\bibnamefont {Perche}},\ }\href
	{https://doi.org/10.1016/j.physletb.2021.136312} {\bibfield  {journal}
		{\bibinfo  {journal} {Phys. Lett. B}\ }\textbf {\bibinfo {volume} {817}},\
		\bibinfo {pages} {136312} (\bibinfo {year} {2021})}\BibitemShut {NoStop}%
	\bibitem [{\citenamefont {Bokuli\'{c}}\ \emph {et~al.}(2021)\citenamefont
		{Bokuli\'{c}}, \citenamefont {Juri\'{c}},\ and\ \citenamefont
		{Smoli\'{c}}}]{Bokulic2021}%
	\BibitemOpen
	\bibfield  {author} {\bibinfo {author} {\bibfnamefont {A.}~\bibnamefont
			{Bokuli\'{c}}}, \bibinfo {author} {\bibfnamefont {T.}~\bibnamefont
			{Juri\'{c}}},\ and\ \bibinfo {author} {\bibfnamefont {I.}~\bibnamefont
			{Smoli\'{c}}},\ }\href {https://doi.org/10.1103/PhysRevD.103.124059}
	{\bibfield  {journal} {\bibinfo  {journal} {Phys. Rev. D}\ }\textbf {\bibinfo
			{volume} {103}},\ \bibinfo {pages} {124059} (\bibinfo {year}
		{2021})}\BibitemShut {NoStop}%
	\bibitem [{\citenamefont {Eslam~Panah}(2024)}]{EslamPanah2024}%
	\BibitemOpen
	\bibfield  {author} {\bibinfo {author} {\bibfnamefont {B.}~\bibnamefont
			{Eslam~Panah}},\ }\href {https://doi.org/10.1093/ptep/ptae012} {\bibfield
		{journal} {\bibinfo  {journal} {Prog. Theor. Exp. Phys.}\ }\textbf {\bibinfo
			{volume} {2024}},\ \bibinfo {pages} {023E01} (\bibinfo {year}
		{2024})}\BibitemShut {NoStop}%
	\bibitem [{\citenamefont {Ahmed}\ \emph
		{et~al.}(2026{\natexlab{a}})\citenamefont {Ahmed}, \citenamefont
		{Al-Badawi},\ and\ \citenamefont {Sakalli}}]{Ahmed2026a}%
	\BibitemOpen
	\bibfield  {author} {\bibinfo {author} {\bibfnamefont {F.}~\bibnamefont
			{Ahmed}}, \bibinfo {author} {\bibfnamefont {A.}~\bibnamefont {Al-Badawi}},\
		and\ \bibinfo {author} {\bibfnamefont {I.}~\bibnamefont {Sakalli}},\ }\href
	{https://doi.org/10.48550/arXiv.2601.10303} {} (\bibinfo {year}
	{2026}{\natexlab{a}}),\ \Eprint {https://arxiv.org/abs/2601.10303}
	{arXiv:2601.10303 [gr-qc]} \BibitemShut {NoStop}%
	\bibitem [{\citenamefont {Ahmed}\ \emph
		{et~al.}(2026{\natexlab{b}})\citenamefont {Ahmed}, \citenamefont
		{Al-Badawi},\ and\ \citenamefont {Silva}}]{Ahmed2026e}%
	\BibitemOpen
	\bibfield  {author} {\bibinfo {author} {\bibfnamefont {F.}~\bibnamefont
			{Ahmed}}, \bibinfo {author} {\bibfnamefont {A.}~\bibnamefont {Al-Badawi}},\
		and\ \bibinfo {author} {\bibfnamefont {E.~O.}\ \bibnamefont {Silva}},\ }\href
	{https://doi.org/10.48550/arXiv.2602.18488} {} (\bibinfo {year}
	{2026}{\natexlab{b}}),\ \Eprint {https://arxiv.org/abs/2602.18488}
	{arXiv:2602.18488 [physics.gen-ph]} \BibitemShut {NoStop}%
	\bibitem [{\citenamefont {Ahmed}\ \emph
		{et~al.}(2026{\natexlab{c}})\citenamefont {Ahmed}, \citenamefont
		{Al-Badawi},\ and\ \citenamefont {Silva}}]{Ahmed2026c}%
	\BibitemOpen
	\bibfield  {author} {\bibinfo {author} {\bibfnamefont {F.}~\bibnamefont
			{Ahmed}}, \bibinfo {author} {\bibfnamefont {A.}~\bibnamefont {Al-Badawi}},\
		and\ \bibinfo {author} {\bibfnamefont {E.~O.}\ \bibnamefont {Silva}},\ }\href
	{https://doi.org/10.48550/arXiv.2602.07806} {} (\bibinfo {year}
	{2026}{\natexlab{c}}),\ \Eprint {https://arxiv.org/abs/2602.07806}
	{arXiv:2602.07806 [gr-qc]} \BibitemShut {NoStop}%
	\bibitem [{\citenamefont {Ahmed}\ \emph
		{et~al.}(2026{\natexlab{d}})\citenamefont {Ahmed}, \citenamefont
		{Al-Badawi},\ and\ \citenamefont {Silva}}]{Ahmed2026d}%
	\BibitemOpen
	\bibfield  {author} {\bibinfo {author} {\bibfnamefont {F.}~\bibnamefont
			{Ahmed}}, \bibinfo {author} {\bibfnamefont {A.}~\bibnamefont {Al-Badawi}},\
		and\ \bibinfo {author} {\bibfnamefont {E.~O.}\ \bibnamefont {Silva}},\ }\href
	{https://doi.org/10.48550/arXiv.2602.02116} {} (\bibinfo {year}
	{2026}{\natexlab{d}}),\ \Eprint {https://arxiv.org/abs/2602.02116}
	{arXiv:2602.02116 [gr-qc]} \BibitemShut {NoStop}%
	\bibitem [{\citenamefont {Bokuli\'c}\ and\ \citenamefont
		{Herdeiro}(2025)}]{Bokulic2025}%
	\BibitemOpen
	\bibfield  {author} {\bibinfo {author} {\bibfnamefont {A.}~\bibnamefont
			{Bokuli\'c}}\ and\ \bibinfo {author} {\bibfnamefont {C.~A.~R.}\ \bibnamefont
			{Herdeiro}},\ }\href {https://doi.org/10.1103/PhysRevD.111.064046} {\bibfield
		{journal} {\bibinfo  {journal} {Phys. Rev. D}\ }\textbf {\bibinfo {volume}
			{111}},\ \bibinfo {pages} {064046} (\bibinfo {year} {2025})}\BibitemShut
	{NoStop}%
	\bibitem [{\citenamefont {Kalb}\ and\ \citenamefont
		{Ramond}(1974)}]{KalbRamond1974}%
	\BibitemOpen
	\bibfield  {author} {\bibinfo {author} {\bibfnamefont {M.}~\bibnamefont
			{Kalb}}\ and\ \bibinfo {author} {\bibfnamefont {P.}~\bibnamefont {Ramond}},\
	}\href {https://doi.org/10.1103/PhysRevD.9.2273} {\bibfield  {journal}
		{\bibinfo  {journal} {Phys. Rev. D}\ }\textbf {\bibinfo {volume} {9}},\
		\bibinfo {pages} {2273} (\bibinfo {year} {1974})}\BibitemShut {NoStop}%
	\bibitem [{\citenamefont {Kosteleck\'{y}}\ and\ \citenamefont
		{Samuel}(1989)}]{KosteleckySamuel1989}%
	\BibitemOpen
	\bibfield  {author} {\bibinfo {author} {\bibfnamefont {V.~A.}\ \bibnamefont
			{Kosteleck\'{y}}}\ and\ \bibinfo {author} {\bibfnamefont {S.}~\bibnamefont
			{Samuel}},\ }\href {https://doi.org/10.1103/PhysRevD.39.683} {\bibfield
		{journal} {\bibinfo  {journal} {Phys. Rev. D}\ }\textbf {\bibinfo {volume}
			{39}},\ \bibinfo {pages} {683} (\bibinfo {year} {1989})}\BibitemShut
	{NoStop}%
	\bibitem [{\citenamefont {Yang}\ \emph {et~al.}(2023)\citenamefont {Yang},
		\citenamefont {Chen}, \citenamefont {Duan},\ and\ \citenamefont
		{Zhao}}]{Yang2023}%
	\BibitemOpen
	\bibfield  {author} {\bibinfo {author} {\bibfnamefont {K.}~\bibnamefont
			{Yang}}, \bibinfo {author} {\bibfnamefont {Y.-Z.}\ \bibnamefont {Chen}},
		\bibinfo {author} {\bibfnamefont {Z.-Q.}\ \bibnamefont {Duan}},\ and\
		\bibinfo {author} {\bibfnamefont {J.-Y.}\ \bibnamefont {Zhao}},\ }\href
	{https://doi.org/10.1103/PhysRevD.108.124004} {\bibfield  {journal} {\bibinfo
			{journal} {Phys. Rev. D}\ }\textbf {\bibinfo {volume} {108}},\ \bibinfo
		{pages} {124004} (\bibinfo {year} {2023})}\BibitemShut {NoStop}%
	\bibitem [{\citenamefont {Lessa}\ \emph {et~al.}(2020)\citenamefont {Lessa},
		\citenamefont {Silva}, \citenamefont {Maluf},\ and\ \citenamefont
		{Almeida}}]{Lessa2020}%
	\BibitemOpen
	\bibfield  {author} {\bibinfo {author} {\bibfnamefont {L.~A.}\ \bibnamefont
			{Lessa}}, \bibinfo {author} {\bibfnamefont {J.~E.~G.}\ \bibnamefont {Silva}},
		\bibinfo {author} {\bibfnamefont {R.~V.}\ \bibnamefont {Maluf}},\ and\
		\bibinfo {author} {\bibfnamefont {C.~A.~S.}\ \bibnamefont {Almeida}},\ }\href
	{https://doi.org/10.1140/epjc/s10052-020-7902-1} {\bibfield  {journal}
		{\bibinfo  {journal} {Eur. Phys. J. C}\ }\textbf {\bibinfo {volume} {80}},\
		\bibinfo {pages} {335} (\bibinfo {year} {2020})}\BibitemShut {NoStop}%
	\bibitem [{\citenamefont {Junior}\ \emph {et~al.}(2024)\citenamefont {Junior},
		\citenamefont {Junior}, \citenamefont {Lobo}, \citenamefont {Rodrigues},
		\citenamefont {Rubiera-Garcia}, \citenamefont {da~Silva},\ and\ \citenamefont
		{Vieira}}]{Junior2024}%
	\BibitemOpen
	\bibfield  {author} {\bibinfo {author} {\bibfnamefont {E.~L.~B.}\
			\bibnamefont {Junior}}, \bibinfo {author} {\bibfnamefont {J.~T. S.~S.}\
			\bibnamefont {Junior}}, \bibinfo {author} {\bibfnamefont {F.~S.~N.}\
			\bibnamefont {Lobo}}, \bibinfo {author} {\bibfnamefont {M.~E.}\ \bibnamefont
			{Rodrigues}}, \bibinfo {author} {\bibfnamefont {D.}~\bibnamefont
			{Rubiera-Garcia}}, \bibinfo {author} {\bibfnamefont {L.~F.~D.}\ \bibnamefont
			{da~Silva}},\ and\ \bibinfo {author} {\bibfnamefont {H.~A.}\ \bibnamefont
			{Vieira}},\ }\href {https://doi.org/10.1103/PhysRevD.110.024077} {\bibfield
		{journal} {\bibinfo  {journal} {Phys. Rev. D}\ }\textbf {\bibinfo {volume}
			{110}},\ \bibinfo {pages} {024077} (\bibinfo {year} {2024})}\BibitemShut
	{NoStop}%
	\bibitem [{\citenamefont {Maluf}\ and\ \citenamefont
		{Neves}(2024)}]{Filho2024}%
	\BibitemOpen
	\bibfield  {author} {\bibinfo {author} {\bibfnamefont {R.~V.}\ \bibnamefont
			{Maluf}}\ and\ \bibinfo {author} {\bibfnamefont {J.~C.~S.}\ \bibnamefont
			{Neves}},\ }\href {https://doi.org/10.1140/epjc/s10052-024-13188-5}
	{\bibfield  {journal} {\bibinfo  {journal} {Eur. Phys. J. C}\ }\textbf
		{\bibinfo {volume} {84}},\ \bibinfo {pages} {827} (\bibinfo {year}
		{2024})}\BibitemShut {NoStop}%
	\bibitem [{\citenamefont {Liu}\ \emph {et~al.}(2024)\citenamefont {Liu} \emph
		{et~al.}}]{Liu2024constraints}%
	\BibitemOpen
	\bibfield  {author} {\bibinfo {author} {\bibfnamefont {C.}~\bibnamefont
			{Liu}} \emph {et~al.},\ }\href
	{https://doi.org/10.1140/epjc/s10052-024-13619-3} {\bibfield  {journal}
		{\bibinfo  {journal} {Eur. Phys. J. C}\ }\textbf {\bibinfo {volume} {84}},\
		\bibinfo {pages} {1185} (\bibinfo {year} {2024})}\BibitemShut {NoStop}%
	\bibitem [{\citenamefont {Letelier}(1979)}]{Letelier1979}%
	\BibitemOpen
	\bibfield  {author} {\bibinfo {author} {\bibfnamefont {P.~S.}\ \bibnamefont
			{Letelier}},\ }\href {https://doi.org/10.1103/PhysRevD.20.1294} {\bibfield
		{journal} {\bibinfo  {journal} {Phys. Rev. D}\ }\textbf {\bibinfo {volume}
			{20}},\ \bibinfo {pages} {1294} (\bibinfo {year} {1979})}\BibitemShut
	{NoStop}%
	\bibitem [{\citenamefont {Toledo}\ and\ \citenamefont
		{Bezerra}(2019)}]{Toledo2019}%
	\BibitemOpen
	\bibfield  {author} {\bibinfo {author} {\bibfnamefont {J.~M.}\ \bibnamefont
			{Toledo}}\ and\ \bibinfo {author} {\bibfnamefont {V.~B.}\ \bibnamefont
			{Bezerra}},\ }\href {https://doi.org/10.1142/S0218271819500238} {\bibfield
		{journal} {\bibinfo  {journal} {Int. J. Mod. Phys. D}\ }\textbf {\bibinfo
			{volume} {28}},\ \bibinfo {pages} {1950023} (\bibinfo {year}
		{2019})}\BibitemShut {NoStop}%
	\bibitem [{\citenamefont {Chabab}\ and\ \citenamefont
		{Iraoui}(2020)}]{Chabab2020}%
	\BibitemOpen
	\bibfield  {author} {\bibinfo {author} {\bibfnamefont {M.}~\bibnamefont
			{Chabab}}\ and\ \bibinfo {author} {\bibfnamefont {S.}~\bibnamefont
			{Iraoui}},\ }\href {https://doi.org/10.1007/s10714-020-02729-4} {\bibfield
		{journal} {\bibinfo  {journal} {General Relativity and Gravitation}\ }\textbf
		{\bibinfo {volume} {52}},\ \bibinfo {pages} {75} (\bibinfo {year}
		{2020})}\BibitemShut {NoStop}%
	\bibitem [{\citenamefont {Ahmed}\ \emph {et~al.}(2022)\citenamefont {Ahmed},
		\citenamefont {Singh},\ and\ \citenamefont {Ghosh}}]{Belhaj2022}%
	\BibitemOpen
	\bibfield  {author} {\bibinfo {author} {\bibfnamefont {F.}~\bibnamefont
			{Ahmed}}, \bibinfo {author} {\bibfnamefont {D.~V.}\ \bibnamefont {Singh}},\
		and\ \bibinfo {author} {\bibfnamefont {S.~G.}\ \bibnamefont {Ghosh}},\ }\href
	{https://doi.org/10.1007/s10714-022-02906-7} {\bibfield  {journal} {\bibinfo
			{journal} {General Relativity and Gravitation}\ }\textbf {\bibinfo {volume}
			{54}},\ \bibinfo {pages} {21} (\bibinfo {year} {2022})}\BibitemShut {NoStop}%
	\bibitem [{\citenamefont {Ahmed}\ \emph
		{et~al.}(2026{\natexlab{e}})\citenamefont {Ahmed}, \citenamefont
		{Al-Badawi},\ and\ \citenamefont {Silva}}]{Ahmed2026b}%
	\BibitemOpen
	\bibfield  {author} {\bibinfo {author} {\bibfnamefont {F.}~\bibnamefont
			{Ahmed}}, \bibinfo {author} {\bibfnamefont {A.}~\bibnamefont {Al-Badawi}},\
		and\ \bibinfo {author} {\bibfnamefont {E.~O.}\ \bibnamefont {Silva}},\ }\href
	{https://doi.org/10.48550/arXiv.2602.15570} {} (\bibinfo {year}
	{2026}{\natexlab{e}}),\ \Eprint {https://arxiv.org/abs/2602.15570}
	{arXiv:2602.15570 [gr-qc]} \BibitemShut {NoStop}%
	\bibitem [{\citenamefont {Perlick}\ and\ \citenamefont
		{Tsupko}(2022)}]{Volker2022}%
	\BibitemOpen
	\bibfield  {author} {\bibinfo {author} {\bibfnamefont {V.}~\bibnamefont
			{Perlick}}\ and\ \bibinfo {author} {\bibfnamefont {O.~Y.}\ \bibnamefont
			{Tsupko}},\ }\href {https://doi.org/10.1016/j.physrep.2021.10.004} {\bibfield
		{journal} {\bibinfo  {journal} {Phys. Rep.}\ }\textbf {\bibinfo {volume}
			{947}},\ \bibinfo {pages} {1} (\bibinfo {year} {2022})}\BibitemShut {NoStop}%
	\bibitem [{\citenamefont {Bekenstein}(1972)}]{Bekestein1972}%
	\BibitemOpen
	\bibfield  {author} {\bibinfo {author} {\bibfnamefont {J.~D.}\ \bibnamefont
			{Bekenstein}},\ }\href {https://doi.org/https://doi.org/10.1007/BF02762768}
	{\bibfield  {journal} {\bibinfo  {journal} {Lett. Nuovo Cim.}\ }\textbf
		{\bibinfo {volume} {4}},\ \bibinfo {pages} {737} (\bibinfo {year}
		{1972})}\BibitemShut {NoStop}%
	\bibitem [{\citenamefont {Bekenstein}(1973)}]{Bekestein1973}%
	\BibitemOpen
	\bibfield  {author} {\bibinfo {author} {\bibfnamefont {J.~D.}\ \bibnamefont
			{Bekenstein}},\ }\href {https://doi.org/10.1103/PhysRevD.7.2333} {\bibfield
		{journal} {\bibinfo  {journal} {Phys. Rev. D}\ }\textbf {\bibinfo {volume}
			{7}},\ \bibinfo {pages} {2333} (\bibinfo {year} {1973})}\BibitemShut
	{NoStop}%
	\bibitem [{\citenamefont {Bardeen}\ \emph {et~al.}(1973)\citenamefont
		{Bardeen}, \citenamefont {Carter},\ and\ \citenamefont
		{Hawking}}]{Bardeen1973}%
	\BibitemOpen
	\bibfield  {author} {\bibinfo {author} {\bibfnamefont {J.~M.}\ \bibnamefont
			{Bardeen}}, \bibinfo {author} {\bibfnamefont {B.}~\bibnamefont {Carter}},\
		and\ \bibinfo {author} {\bibfnamefont {S.~W.}\ \bibnamefont {Hawking}},\
	}\href {https://doi.org/10.1007/BF01645742} {\bibfield  {journal} {\bibinfo
			{journal} {Commun. Math. Phys.}\ }\textbf {\bibinfo {volume} {31}},\ \bibinfo
		{pages} {161} (\bibinfo {year} {1973})}\BibitemShut {NoStop}%
	\bibitem [{\citenamefont {Hawking}(1975)}]{Hawking1975}%
	\BibitemOpen
	\bibfield  {author} {\bibinfo {author} {\bibfnamefont {S.~W.}\ \bibnamefont
			{Hawking}},\ }\href {https://doi.org/10.1007/BF02345020} {\bibfield
		{journal} {\bibinfo  {journal} {Commun. Math. Phys.}\ }\textbf {\bibinfo
			{volume} {43}},\ \bibinfo {pages} {199} (\bibinfo {year} {1975})}\BibitemShut
	{NoStop}%
	\bibitem [{\citenamefont {Wald}(1993)}]{Wald1993}%
	\BibitemOpen
	\bibfield  {author} {\bibinfo {author} {\bibfnamefont {R.~M.}\ \bibnamefont
			{Wald}},\ }\href {https://doi.org/10.1103/PhysRevD.48.R3427} {\bibfield
		{journal} {\bibinfo  {journal} {Phys. Rev. D}\ }\textbf {\bibinfo {volume}
			{48}},\ \bibinfo {pages} {3427} (\bibinfo {year} {1993})}\BibitemShut
	{NoStop}%
	\bibitem [{\citenamefont {Iyer}\ and\ \citenamefont
		{Wald}(1994)}]{IyerWald1994}%
	\BibitemOpen
	\bibfield  {author} {\bibinfo {author} {\bibfnamefont {V.}~\bibnamefont
			{Iyer}}\ and\ \bibinfo {author} {\bibfnamefont {R.~M.}\ \bibnamefont
			{Wald}},\ }\href {https://doi.org/10.1103/PhysRevD.50.846} {\bibfield
		{journal} {\bibinfo  {journal} {Phys. Rev. D}\ }\textbf {\bibinfo {volume}
			{50}},\ \bibinfo {pages} {846} (\bibinfo {year} {1994})}\BibitemShut
	{NoStop}%
	\bibitem [{\citenamefont {Page}(1976)}]{Page1976}%
	\BibitemOpen
	\bibfield  {author} {\bibinfo {author} {\bibfnamefont {D.~N.}\ \bibnamefont
			{Page}},\ }\href {https://doi.org/10.1103/PhysRevD.13.198} {\bibfield
		{journal} {\bibinfo  {journal} {Phys. Rev. D}\ }\textbf {\bibinfo {volume}
			{13}},\ \bibinfo {pages} {198} (\bibinfo {year} {1976})}\BibitemShut
	{NoStop}%
	\bibitem [{\citenamefont {Gray}\ \emph {et~al.}(2016)\citenamefont {Gray},
		\citenamefont {Schuster}, \citenamefont {Van-Brunt},\ and\ \citenamefont
		{Visser}}]{Gray2016}%
	\BibitemOpen
	\bibfield  {author} {\bibinfo {author} {\bibfnamefont {F.}~\bibnamefont
			{Gray}}, \bibinfo {author} {\bibfnamefont {S.}~\bibnamefont {Schuster}},
		\bibinfo {author} {\bibfnamefont {A.}~\bibnamefont {Van-Brunt}},\ and\
		\bibinfo {author} {\bibfnamefont {M.}~\bibnamefont {Visser}},\ }\href
	{https://doi.org/10.1088/0264-9381/33/11/115003} {\bibfield  {journal}
		{\bibinfo  {journal} {Class. Quantum Grav.}\ }\textbf {\bibinfo {volume}
			{33}},\ \bibinfo {pages} {115003} (\bibinfo {year} {2016})}\BibitemShut
	{NoStop}%
	\bibitem [{\citenamefont {Javed}\ \emph {et~al.}(2019)\citenamefont {Javed},
		\citenamefont {Abbas},\ and\ \citenamefont {\"{O}vg\"{u}n}}]{Javed2019}%
	\BibitemOpen
	\bibfield  {author} {\bibinfo {author} {\bibfnamefont {W.}~\bibnamefont
			{Javed}}, \bibinfo {author} {\bibfnamefont {J.}~\bibnamefont {Abbas}},\ and\
		\bibinfo {author} {\bibfnamefont {A.}~\bibnamefont {\"{O}vg\"{u}n}},\ }\href
	{https://doi.org/10.1103/PhysRevD.100.044052} {\bibfield  {journal} {\bibinfo
			{journal} {Phys. Rev. D}\ }\textbf {\bibinfo {volume} {100}},\ \bibinfo
		{pages} {044052} (\bibinfo {year} {2019})}\BibitemShut {NoStop}%
	\bibitem [{\citenamefont {Misner}\ \emph {et~al.}(1973)\citenamefont {Misner},
		\citenamefont {Thorne},\ and\ \citenamefont {Wheeler}}]{Misner1973}%
	\BibitemOpen
	\bibfield  {author} {\bibinfo {author} {\bibfnamefont {C.~W.}\ \bibnamefont
			{Misner}}, \bibinfo {author} {\bibfnamefont {K.~S.}\ \bibnamefont {Thorne}},\
		and\ \bibinfo {author} {\bibfnamefont {J.~A.}\ \bibnamefont {Wheeler}},\
	}\href@noop {} {\emph {\bibinfo {title} {Gravitation}}}\ (\bibinfo
	{publisher} {W. H. Freeman},\ \bibinfo {address} {San Francisco},\ \bibinfo
	{year} {1973})\BibitemShut {NoStop}%
	\bibitem [{\citenamefont {Mashhoon}(1973)}]{Mashhoon1973}%
	\BibitemOpen
	\bibfield  {author} {\bibinfo {author} {\bibfnamefont {B.}~\bibnamefont
			{Mashhoon}},\ }\href {https://doi.org/10.1103/PhysRevD.7.2807} {\bibfield
		{journal} {\bibinfo  {journal} {Phys. Rev. D}\ }\textbf {\bibinfo {volume}
			{7}},\ \bibinfo {pages} {2807} (\bibinfo {year} {1973})}\BibitemShut
	{NoStop}%
	\bibitem [{\citenamefont {Wei}\ and\ \citenamefont {Liu}(2013)}]{Wei2013}%
	\BibitemOpen
	\bibfield  {author} {\bibinfo {author} {\bibfnamefont {S.-W.}\ \bibnamefont
			{Wei}}\ and\ \bibinfo {author} {\bibfnamefont {Y.-X.}\ \bibnamefont {Liu}},\
	}\href {https://doi.org/10.1088/1475-7516/2013/11/063} {\bibfield  {journal}
		{\bibinfo  {journal} {JCAP}\ }\textbf {\bibinfo {volume} {2013}}\bibinfo
		{number} { (11)},\ \bibinfo {pages} {063}}\BibitemShut {NoStop}%
\end{thebibliography}
\end{document}